\documentclass[12pt]{article}
\usepackage[space]{cite}
\usepackage{amssymb,graphicx}
\usepackage{epstopdf}
\usepackage{amsmath,amsfonts}
\usepackage{epsfig}
\usepackage{textcomp}
\usepackage{hyperref}

\def\e{{\rm e}}
\def\al{\alpha}
\def\tbeta{\tilde{\beta}}

\def\d{\partial}
\def\l{\left(}
\def\r{\right)}

\def\t0{\tilde{0}}
\def\ta{\tilde{a}}
\def\tb{\tilde{b}}
\def\tc{\tilde{c}}

\newcommand{\be}{\begin{equation}}
\newcommand{\ee}{\end{equation}}
\newcommand{\bea}{\begin{eqnarray}}
\newcommand{\eea}{\end{eqnarray}}
\newcommand{\bg}{\begin{gather}}
\newcommand{\eg}{\end{gather}}
\newcommand{\bseq}{\begin{subequations}}
\newcommand{\eseq}{\end{subequations}}

\newcommand{\talpha}{\tilde{\alpha}}

\begin{document}
\vspace{10pt}
\begin{center}
  {\LARGE \bf On the stability of self-accelerating Universe in modified gravity with dynamical torsion} \\
\vspace{20pt}
V.~Nikiforova$^a$\\
\vspace{15pt}
  $^a$\textit{Institute for Nuclear Research of
         the Russian Academy of Sciences,\\  60th October Anniversary
  Prospect 7a, 117312 Moscow, Russia
  }\\
    \end{center}
    \vspace{5pt}

\begin{abstract}
We consider the model of modified gravity with dynamical torsion. This model was previously found to have promising stability properties about various backgrounds. Here we study the stability of linear perturbations about the self-accelerating solution. We apply the $(3+1)$-decomposition and consider the scalar sector of perturbations. We find that the number of degrees of freedom is equal to 2, which is the same as in Minkowski background. However, there is at least one instability in the scalar sector, if the value of background torsion is large enough. This does not rule out the possibility of stable self-acceleration with torsion of the order of the effective cosmological constant.
\end{abstract}


\section{Introduction}
Cosmological observations show that the expansion of our Universe is accelerating. The mechanism of acceleration is, however, not understood. One way of solving the acceleration problem is the IR-modification of gravity. 

Many IR-modified gravitational theories have been proposed, see reviews \cite{1, 2, 3, 4, 5, 6}. However, the self-accelerating solutions are often unstable because of the ghost and/or gradient instabilities in the spectrum of the linearized perturbations. 

We focus here on gravities with dynamical torsion, which are promising candidates for the consistent infrared modified gravity. Gravity with dynamical torsion is a natural generalization of General Relativity which treats the connection and vierbein as independent dynamical variables. The torsion is capable of propagating due to the terms in the Lagrangian quadratic in torsion and curvature. These theories are often considered in the framework of Poincar\'e gauge gravities \cite{book1, book2, book3, HLecture}.

The spectrum of linearized perturbations in gravities with dynamical torsion contains additional degrees of freedom as compared to General Relativity. Not all theories from this class are stable about Minkowski background at the linearized level. The stability about Minkowski background was thoroughly investigated in Refs.~\cite{HS-1, HS-2, HS-3, Sezgin} where stable models were identified.

Self-accelerating solutions, without explicit cosmological constant term in the action, were found in various gravities with dynamical torsion in Refs.~\cite{H2, 42, 49, 44, 50, 47, 51, 46, 48, last}. The open question is the stability of self-accelerating solutions in these theories, at least at the linearized level. It is the question we address in the present paper.

Let us mention that there is another approach to make torsion propagating, so-called f(T) gravities. These theories, as well as the  cosmological solutions and perturbations about the cosmological solutions in these theories, are widely investigated, for example, in \cite{fT1, fT2, fT3, fT4, fT5} and references therein. 

We consider a particular model studied in Refs.~\cite{45, 33, 34, last}. This model has nice stability properties. Namely, it was shown that ghosts, gradient instabilities and tachyons are absent in the Minkowski background, de Sitter and anti-de Sitter spaces and arbitrary torsionless Einstein backgrounds of sufficiently small curvature \cite{HS-3, 33, 34}. The model also admits self-accelerating solution, without explicit cosmological constant term in the action \cite{last}, with self-acceleration due to the non-trivial connection. 

In this paper we study linear perturbations about the self-accelerating solution found in Ref.~\cite{last}. Known examples show that the most dangerous in this regard is the scalar sector of perturbations. Therefore, we focus in this paper on the scalar sector. 

The first issue to worry about is the possible existence of the Bouleware--Deser modes \cite{BD}.  In other models it often happens that the Minkowski background is stable, but extra modes appear in curved backgrounds. These modes usually have wrong sign kinetic terms. However, this pathology is not necessary present: one famous counterexample is dRGT model \cite{dRGT, no_ghosts-1, no_ghosts-2}.  Other dangerous features are the ghost and gradient instabilities.

This paper is organized as follows. In Section 2 we present the Lagrangian, write the field equations and recall the earlier results. In Section 3 we derive the generalized Bianchi identities. We introduce conformal time and perform conformal transformation in Section 4. In Section 5 we make $(3+1)$-decomposition of perturbations. Section 6 contains our main results. We study the scalar perturbations about the self-accelerating solution. We show that there are no Bouleware--Deser mode in the scalar sector. The number of dynamical degrees of freedom is the same as in Minkowski background and is equal to 2. However, we show that at least one of these degrees of freedom is exponentially growing if the value of background torsion is large enough. We conclude in Section 7.

\section{The Model} \label{sec:1}
We make use of the tetrad formalism and consider vierbein and connection as independent fields. Following the notations of Refs. \cite{45, 33, 34, last}, we denote the vierbein by $e^i_\mu$ and connection by $A_{ij\mu} = -A_{ji\mu}$, where $\mu = (0,1,2,3)$ is the space-time index, and $i,j = (0,1,2,3)$ are the tangent space indices. In tangent space basis the indices are raised and lowered using the Minkowski metric $\eta_{ij}$, so we do not distinguish upper and lower tangent space indices in what follows, if this does not lead to an ambiguity. The signature of metric is $(-, +, +, +)$.

The action of the model is
\be  S= \int~d^4x~eL  \; , \;\;\; L= \frac{3}{2} ( \tilde{\alpha} F -
\alpha R) +
 c_3 F^{ij}F_{ij} + c_4 F^{ij}F_{ji} + c_5 F^2 + c_6 (\epsilon \cdot F)^2 \; , \label{all_2} \ee
where $ \alpha, \tilde{\alpha}, c_3, c_4, c_5, c_6$ are coupling constants, 
 $ e \equiv det(e^i_\mu) ; $
$F_{ijkl}$ is the curvature tensor constructed with the connection $A_{ij\mu}$,
\be
F_{ijkl} = e^\mu_k e^\nu_l ( \partial_\mu A_{ij\nu} - \partial_\nu A_{ij\mu} + A_{im\mu}A_{m j\nu} - A_{jm\mu}A_{m i\nu} ) \; ; \nonumber
\ee
\be F_{ij}=\eta^{kl}F_{ikjl}\;,\;\; F=\eta^{ij}F_{ij} \; , \;\; \epsilon \cdot F \equiv \epsilon^{ijkl}F_{ijkl} \nonumber \; ;\ee
$\epsilon_{ijkl}$ is the Levi-Civita symbol defined in such a way that $\epsilon^{0123} = - \epsilon_{0123} = 1$;
$R_{ijkl}$ is the Riemannian curvature tensor,
\be R_{ijkl}= e^\mu_k e^\nu_l ( \partial_\mu \omega_{ij\nu} - \partial_\nu
\omega_{ij\mu} + \omega_{im\mu}\omega_{mj\nu} - \omega_{jm\mu}\omega_{mi\nu} )\; ; \nonumber
\ee \be R_{ij}=\eta^{kl}R_{ikjl}\;,\;\; R=\eta^{ij}R_{ij}\; , \nonumber \ee
where $\omega_{ij\mu}$ is the Riemannian spin-connection. It is expressed in terms of the vierbein as follows:
\be
 \omega_{ij\mu} \equiv \omega_{ijk}e^k_\mu =\frac{1}{2}( C_{ijk}-C_{jik} - C_{kij} )e^k_\mu \; ,  \nonumber
\ee
where
\be
C_ {ijk} = e_j^\mu e_k^\nu (\d_\mu e_{i\nu} - \d_\nu e_{i\mu}) \; . \nonumber
\ee
Its relation to Christoffel symbols is
\be
\omega_{ij\mu}=-\Gamma^\nu_{\mu\lambda}e_i^\lambda e_{j\nu} - e_{j\nu}\d_\mu e_i^\nu \;.   \nonumber
\ee
The connection $A_{ij\mu}$ can be represented as a sum
\be
 A_{ij\mu}=\omega_{ij\mu}+K_{ij\mu}\; , \nonumber
\ee
where $K_{ij\mu}$ is the contorsion tensor.

We impose the following conditions on the couplings:
\begin{align}
& c_3 \neq c_4 \label{all_340} \;, \\
& c_3+c_4=-3c_5 \;, \label{all_19} \\
&\alpha < 0, \;\;\;\; \tilde{\alpha} > 0, \;\;\;\; c_5 < 0, \;\;\;\; c_6 > 0  \label{all_13} \; . \end{align}
The reason for imposing the condition \eqref{all_340} will become clear in Sec.~\ref{SSr}, while the conditions \eqref{all_19} and \eqref{all_13} ensure that there are only healthy degrees of freedom in the Minkowski background \cite{HS-1, HS-2, HS-3, Sezgin, 45, 33}. It was shown in Ref.~\cite{33} that the strength of the gravitational interaction between the energy-momentum tensors is govered by the parameters $\alpha$ and $\talpha$. We assume that $|\al| \sim \talpha$, and hence \be  |\al| \sim \talpha \sim M_{Pl}^2 \;. \label{all_328}  \ee

In Ref. \cite{34} it was found that there are three propagating modes at the linearized level in the Minkowski background: the massless spin-2 mode, the massive spin-2 mode with mass
\be m^2=\frac{\talpha(\tilde{\alpha}-\alpha)}{2\alpha c_5} \label{all_80} \ee
 and the massive spin-0 mode with mass \be m^2_0=\frac{\tilde{\alpha}}{16c_6} \;. \label{all_90} \ee There are no ghosts or tachyons in the Minkowski background. It was also shown that in the theory equipped with the cosmological constant, the perturbations are healthy in maximally symmetric backgrounds, as well as in torsionless Einstein backgrounds of sufficiently small curvature.

There are two sets of field equations in our model. One consists of the gravitational field equations obtained by varying the action with respect to vierbein,
\begin{align}
\hat{{\cal G}}_{ij} \equiv &  \frac{3}{2} \tilde{\alpha} \left( F_{ij} - \frac{1}{2} \eta_{ij}F  \right) -
\frac{3}{2} \alpha \left( R_{ij} - \frac{1}{2} \eta_{ij}R  \right) + c_3 \left( F_{ki} F_{kj} + F_{kl} F_{kilj} \right)
\nonumber\\
& + c_4 \left( F_{ik}
F_{kj} + F_{lk} F_{kilj} \right) + 2 c_5 F F_{ij} \nonumber \\
& + 2 c_6 \epsilon_{klmi} F_{klmj} (\epsilon\cdot F) -
\frac{1}{2}\eta_{ij} L^{(2)}  = 0 \;,\label{all-1}
\end{align}
where
\be L^{(2)}  = c_3 F_{ij}F_{ij} + c_4 F_{ij}F_{ji} + c_5 F^2 + c_6 (\epsilon \cdot F)^2 \nonumber  \ee
is the part in the Lagrangian which is bilinear in $F_{ijkl}$.
Another set of equations is obtained by varying the action with respect to the connection $A_{ij\mu}$,
\begin{align}
\hat{{\cal T}}_{ijk} \equiv  & H_{ijk}  + \left\{ \left[\eta_{ik} \left( D_m P_{jm} - \frac{2}{3}
D_j P\right) - D_i P_{jk}\right]  - \left[\eta_{jk} \left( D_m P_{im} -
\frac{2}{3} D_i P\right) - D_j P_{ik} \right]\right\}
\nonumber\\
& +4 c_6 \epsilon_{ijkm} D_m (\epsilon\cdot F) + S_{ijk} = 0 \;,
\label{all-2}
\end{align}
where $P_{ij}$ and $P$ are defined as follows:
\begin{align*}   & P_{ij} = c_3F_{ij} + c_4 F_{ji}\;, \nonumber \\ & P=\eta^{ij}P_{ij} \;;   \end{align*}
the covariant derivative $D_i$ involves the connection $A_{ij\mu}$,
\be  D_iB_j  \equiv  e^\mu_iD_\mu B_j = e^\mu_i( \partial_\mu B_{j} - A_{ lj\mu}B_{l} ) \;; \nonumber \ee
$S_{ijk}$ is defined as follows: \be S_{ijk} = \frac{2}{3\talpha} H_{mnk} \l \eta_{im} P_{jn}
- \eta_{jm} P_{in} -\frac{2}{3} \eta_{im}\eta_{jn} P + 2c_6
\epsilon_{ijmn} (\epsilon \cdot F) \r \;,  \nonumber \ee  and $H_{ijk}$ is written in terms of
contorsion: \be \frac{2}{3\talpha} H_{ijk} = K_{ikj} -
K_{jki} - K_{ill} \eta_{jk} + K_{jll} \eta_{ik} \;. \label{all_41} \ee The emergence of
this object is clarified in Appendix A.

Note that because of invariance of the action under space-time gauge transformations and infinitesimal local Lorentz transformations, not all of the field equations are independent.

In Ref.~\cite{33} it was shown that the model \eqref{all_2} admits a self-accelerating cosmological solution with spatially flat metric,
\be  e^{\t0}_0 = 1 \;,\;\;\;\; e^{\ta }_b = e^{\lambda t} \delta^{\ta}_b
 \;,\quad
A_{\t0 \ta \tb}=f\delta_{\ta\tb}  \;, \quad
A_{\ta\tb\tc}=g\varepsilon_{\ta\tb\tc}\;,  \label{all_81}  \ee
with time-independent $\lambda$, $f$ and $g$, where $a, \tilde{a}=(1,2,3)$, tilde denotes tangent space indices, and space-time indices do not have tilde. The parameters $\lambda$, $f$, $g$ and $\alpha$, $\talpha$, $c_5$, $c_6$ are related as follows,
\begin{align}
&c_6=\frac{\talpha\lambda (\talpha f+\al \lambda)}{16(\lambda^2-4f^2)(\talpha f^2-\talpha\lambda f - 2\al\lambda^2)} \;,  \nonumber
\\
& c_5=\frac{\talpha[2\talpha f^2+\lambda f\talpha+\lambda^2(\talpha-2\al)]}{4\lambda(\lambda+2f)(f^2\talpha-\lambda f\talpha-2\al\lambda^2)}  \;,   \label{all_201}   \\
& g^2=\frac{2\alpha \lambda^2 - \talpha f^2 + \talpha \lambda f}{\talpha} \;. \nonumber
\end{align}
In this paper we consider the case of small $\lambda$ and large enough $f$, \be f \gg \lambda \log^2A \;, \label{all_326} \ee
where $A$ is the initial amplitude of perturbation, see Sec.~\ref{SSr}. In this case, the small value of the effective cosmological constant $\lambda$ is obtained provided that there is a hierarchy between the dimensionless couplings: to the leading order in $\lambda$ \be c_6 = -\lambda \frac{\talpha}{64f^3}  \;, \quad c_5 = \frac{\talpha}{4\lambda f}  \;.  \label{all_301} \ee  In the small-$\lambda$ limit the parameters $f$, $g$ and $\lambda$ are related to the couplings as follows:
\begin{align*}
 & \lambda = \left( -\frac{c_6\talpha^2}{c_5^3} \right)^{1/4}  \;,  \nonumber \\
 & f=-\frac{\talpha^{1/2}}{4(-c_5c_6)^{1/4}}  \;,     \\
 & g=\pm f+O(\lambda) \;. \nonumber
\end{align*}
Note that eq.~\eqref{all_301} implies that the overall scale $|\al|$ enters the action \eqref{all_2} as a pre-factor. This implies that the equations for perturbations, written in terms of $f$, $g$ and $\lambda$, involve the ratio $\talpha/\al$ and not $\talpha$ and $\al$ themselves. The same property holds for dispersion relations.

\section{Generalized Bianchi Identities}  \label{BI}
 Let us obtain the identities relating field equations with each other. The reasoning is similar to that used for deriving the Bianchi identities in General Relativity.

The first identity follows from the invariance under space-time gauge transformations. Making an infinitesimal gauge transformation, $x^{\prime\mu}=x^\mu+\xi^\mu$, we find the variations of $e^i_{\mu}$ and $A_{ij\mu}$:
\begin{align}
& \delta e^i_\mu = -\xi^\lambda\d_\lambda e^i_\mu-e^i_\lambda\d_\mu \xi^\lambda =  -\xi^\lambda\nabla_\lambda e^i_\mu-e^i_\lambda\nabla_\mu \xi^\lambda  \label{all_42} \;, \\
& \delta A_{ij\mu} = - A_{ij\lambda}\nabla_\mu\xi^\lambda - \xi^\lambda\nabla_\lambda A_{ij\mu} \;,  \label{all_43}
\end{align}
where $\nabla_\mu$ is the covariant derivative involving the Christoffel symbols,
\be  \nabla_\mu e^i_\nu  =  \partial_\mu e^i_\nu - \Gamma_{\cdot\mu\nu}^{\lambda}e^{i}_{\lambda}  \;. \nonumber  \ee
Varying the action with respect to $e^i_{\mu}$ and $A_{ij\mu}$ of the form \eqref{all_42}, \eqref{all_43} we find, after integrating by parts, the following identity:
\be
-2 \hat{{\cal G}}_{ik} e_{k \nu}\nabla_\mu  e^{i \nu} - 2 \nabla_\nu (   \hat{{\cal G}}_{ik} e_{k \mu}  e^{i \nu} ) -  e^{k\nu} \nabla_\mu  A_{ij \nu}  \hat{{\cal T}}_{ijk} +  \nabla_\nu (  e^{k\nu} A_{ij \mu} \hat{{\cal T}}_{ijk} ) \equiv 0 \; \label{all_45}
\ee  
(see eqs.~\eqref{all-1} and \eqref{all-2} for definition of $\hat{{\cal G}}_{ij}$ and $\hat{{\cal T}}_{ijk}$). The left hand side of eq.~\eqref{all_45} contains one free index $\mu$.

Another two identities come from the local Lorentz invariance. The model discussed is a gauge theory with Lorentz group as a gauge group. The variations of $e^i_{\mu}$ and $A_{ij\mu}$ under the infinitesimal Lorentz transformation are
\begin{align}
& \delta e_i^\mu =\omega_{ij}e_j^\mu \;, \label{all_105} \\
& \delta A_{ij\mu}= - \partial_\mu\omega_{ij} + \omega_{ip}A_{p j\mu} - \omega_{jp}A_{p i\mu} \;, 
\end{align}
where $\omega_{ij} = -\omega_{ji}$ are the parameters of transformation. The invariance of the action under this transformation gives another identity,
\be
( \hat{{\cal G}}_{ij}- \hat{{\cal G}}_{ji}) +  \nabla_\nu( \hat{{\cal T}}_{ijk} e^{k\nu}) + 2 e^{k\nu}(  \hat{{\cal T}}_{ipk} A_{jp\nu} -  \hat{{\cal T}}_{jpk} A_{ip\nu} ) \equiv 0  \;. \label{all_66}
\ee
The left hand side of eq.~\eqref{all_66} contains two free indices, $i$ and $j$, and is antisymmetric in them. 

As a result, we have two sets of generalized Bianchi identities, which relate the field equations \eqref{all-1} and \eqref{all-2} with each other.

\section{Extracting Conformal Factor from Vierbein} \label{all_52}

The purpose of this paper is to study small perturbations about self-accelerating background \eqref{all_81}. To this end, we define conformal time,
\be
\eta = \int e^{-\lambda t}dt = -\frac{1}{\lambda}e^{-\lambda t} \;. \nonumber
\ee
We work with conformal time in what follows and denote by prime the derivative with respect to it. At this point it is convenient to change the notation. We denote the vierbein of the original theory by $\hat{e}^i_\mu$ and then write,
\be \hat{e}^i_\mu =
\e^{\phi}  e^i_\mu \;,\;\; \hat{e}_j^\nu =
\e^{-\phi}  e_j^\nu \;, \label{all_25} \ee
where
 \be \e^{\phi} = a(\eta) = -\frac{1}{\lambda \eta} \;.  \label{all_401} \ee
Note that
\be  \phi^{\prime} =
\lambda \e^{\phi} \;.
\nonumber \ee
We do not make any scaling of connection, and keep the notation $A_{ij\mu}$ unchanged. Upon the conformal transformation \eqref{all_25}, the background \eqref{all_81} is written in terms of $e^i_{\mu}$ and $A_{ij\mu}$ as follows::
\be  e^i_\mu = \delta^i_\mu
 \;,\quad
   A_{\t0 \ta b}=e^{\phi}f\delta_{\ta b}   \;, \quad
  A_{\ta\tb c}=e^{\phi}g\varepsilon_{\ta\tb c}  \;.  \label{all_141}  \ee
Now, $R_{ij}$ and $R$ transform under the conformal transformation \eqref{all_25} as follows:
\begin{align*}
\hat{R}_{ij} = & \e^{-2\phi} \l R_{ij} -2 e^\mu_i e^\nu_j
\nabla_\mu \nabla_\nu \phi - \eta_{ij} g^{\mu \nu} \nabla_\mu
\nabla_\nu \phi \right.
\\
& \left. + 2  e^\mu_i e^\nu_j \d_\mu \phi \d_\nu \phi -2 \eta_{ij}
g^{\mu \nu} \d_\mu \phi \d_\nu \phi \r \;, 
\\
\hat{R} = &  \e^{-2\phi} \l R - 6  g^{\mu \nu} \nabla_\mu
\nabla_\nu \phi - 6  g^{\mu \nu} \d_\mu \phi \d_\nu \phi \r \;,
\end{align*}
where $\hat{R}_{ij}$ is constructed with $\hat{e}^i_\mu$, while
$R_{ij}$, $\nabla_\mu$ and $g_{\mu \nu} = e^i_\mu e_{i \nu}$ are constructed
with $e^i_\mu$. The original curvature $\hat{F}_{ijkl}$ is \be
\hat{F}_{ijkl} = \e^{-2\phi} F_{ijkl} \;, \nonumber \ee 
where $F_{ijkl}$ is constructed using $e^i_\mu$ and $A_{ij\mu}$.

The gravity equation \eqref{all-1} then reads,
\begin{align}
{\cal G}_{ij} \equiv \; & \frac{3}{2} \talpha \e^{-2\phi} \l F_{ij} - \frac{1}{2}
\eta_{ij}F  \r - \frac{3}{2} \alpha \e^{-2\phi} \l R_{ij} -
\frac{1}{2} \eta_{ij}R  \r
\nonumber\\
&+ \e^{-4\phi}c_3 \l F_{ki} F_{kj} + F_{kl} F_{kilj} \r + \e^{-4\phi}c_4 \l F_{ik}
F_{kj} + F_{lk} F_{kilj} \r + 2 \e^{-4\phi}c_5 F F_{ij}
\nonumber\\
&+ 2 \e^{-4\phi}c_6 \epsilon_{klmi} F_{klmj} (\epsilon\cdot F) -
\e^{-4\phi}\frac{1}{2}\eta_{ij} L^{(2)} + \Delta^{(G)}_{ij}  = 0  \;, \label{all_21}
\end{align}
where
\begin{align*}
\Delta^{(G)}_{ij} = & - \frac{3}{2} \alpha \e^{-2\phi} \left[ -2 e^\mu_i e^\nu_j
\nabla_\mu \nabla_\nu \phi - \eta_{ij} g^{\mu \nu} \nabla_\mu
\nabla_\nu \phi + 2  e^\mu_i e^\nu_j \d_\mu \phi \d_\nu \phi -2
\eta_{ij} g^{\mu \nu} \d_\mu \phi \d_\nu \phi \right.
\\
& \left. - \eta_{ij} \l -3 g^{\mu \nu} \nabla_\mu \nabla_\nu \phi
-3 g^{\mu \nu} \d_\mu \phi \d_\nu \phi \r \right] \;.
\end{align*}
A simple way to obtain the new version of the torsion equation is
to plug the vierbein \eqref{all_25} into the action, and vary with respect to $A_{ij\mu}$. Then the torsion equation is
\begin{align}
{\cal T}_{ijk} \equiv \;& 
\e^{-3\phi}\left\{ \left[\eta_{ik} \l D_m P_{jm} - \frac{2}{3} D_j P\r -
D_i P_{jk}\right]  - \left[\eta_{jk} \l D_m P_{im} - \frac{2}{3}
D_i P\r - D_j P_{ik} \right]\right\}
\nonumber\\
& +4 \e^{-3\phi}c_6 \epsilon_{ijkm} D_m (\epsilon\cdot F) + \e^{-3\phi}S_{ijk} + \e^{-\phi}H_{ijk}   + \Delta^{(T)}_{ijk} = 0 \;,
\label{all_22}
\end{align}
where
\be \Delta^{(T)}_{ijk} = 3\talpha\e^{-\phi}\l    \eta_{ik} e^\mu_j \d_\mu \phi
- \eta_{jk} e^\mu_i \d_\mu \phi \r \;. \nonumber \ee

\section{Perturbations about Self-accelerating Solution: 3+1 Decomposition}

To study the linear perturbations about the self-accelerating background \eqref{all_141}, it is convenient to make the 3-dimensional Fourier decomposition. We use the same notation for the Fourier-transformed variables and replace  spatial derivatives $\d_{\ta} \equiv e_{\ta}^\mu \d_\mu$ with $ik_{\ta}$, where ${\bf k}$ is the 3-dimensional momentum. Since the background \eqref{all_141}  is invariant under spatial rotations, it is natural to use $(3+1)$-decomposition of perturbations. As usual, this means that we decompose any 3-dimensional tensor into its irreducible components with respect to the small group $ SO(2)$ of rotations around the spatial momentum. These irreducible components form sectors with particular helicities: scalar sector (helicity-0), vector sector (helicity-1) and tensor sector (helicity-2). Since the field equations are linear, these sectors can be considered separately.

After conformal transformation, the background vierbein in \eqref{all_141} is trivial, so we do not distinguish space-time and Lorentz indices in $(3+1)$-decomposition of perturbations. These indices are raised and lowered with Minkowski metric, while spatial indices are contracted with Euclidean metric.

\subsection{Perturbation of connection}
The full contorsion tensor can be written as follows,
\be  K_{ijk}={K}_{ijk \; (0)}+k_{ijk}  \;, \nonumber  \ee    where ${K}_{ijk \; (0)}= e^\mu_k{K}_{ij\mu \; (0)}= e^\mu_k({A}_{ij\mu \; (0)} - {\omega}_{ij\mu \; (0)}) = e^\mu_k{A}_{ij\mu \; (0)} $ is the background quantity constructed from \eqref{all_141} and $k_{ijk}$ is the first-order perturbation. Let us decompose the first order contorsion tensor $k_{ijk}$ into its helicity components. 

Tensor $k_{ijk}=-k_{jik}$ contains helicity 0, 1, 2 components only. The helicity-2 components form the tensor sector:
\begin{align*}
k_{0ab} &= - k_{a0b} = \tau_{ab} \;,
\\
k_{abc} &= k_{a} N_{bc} - k_{b} N_{ac} \;,
\end{align*}
where $\tau_{ab}$ and $N_{ab}$ are symmetric, transverse and traceless.
All other components of $k_{ijk}$ vanish in the tensor sector.

The non-vanishing helicity-1 components of contorsion are
\begin{align*}
k_{0a0} & = - k_{a00} = \zeta_{a} \;,
\\
k_{0ab} &= - k_{a0b} = k_{a} \nu_{b} + k_{b} \mu_{a} \;,
\\
k_{ab0} &= k_{a} \kappa_{b} - k_{b} \kappa_{a} \;,
\\
k_{abc} &= k_{a} k_{c} \alpha_{b} - k_{b} k_{c} \alpha_{a} + \eta_{ac} L_{b} -
\eta_{bc} L_{a} \;.
\end{align*}
All 3-vectors here are transverse.
The helicity-0 components form the scalar sector:
\begin{align*}
k_{0a0} & = - k_{a00} = k_{a} \xi \;,
\\
k_{0ab} &= - k_{a0b} = k_{a} k_{b} \chi + \delta_{ab} \sigma +
\epsilon_{abc} k_{c} \rho \;,
\\
k_{ab0} &= \epsilon_{abc} k_{c} \theta \;,
\\
k_{abc} &= \epsilon_{abd} k_{c} k_{d} Q + (k_{a} \epsilon_{bcd} - k_{b}
\epsilon_{acd}) k_{d} u + (k_{a} \delta_{bc} - k_{b} \delta_{ac} ) M \;.
\end{align*}
There are altogether 24 components of tensor $k_{ijk}$. They break up into the 4 tensor components (two components in each of $\tau_{ab}$ and $N_{ab}$), 12 components in the vector sector (two components in each of the six transverse vectors $\zeta_{a}$, $\nu_{a}$, $\mu_{a}$, $\kappa_{a}$, $\alpha_{a}$ and $L_{a}$) and 8 scalar components ($\xi$, $\chi$, $\sigma$, $\rho$, $\theta$, $Q$, $u$ and $M$).

\subsection{Perturbations of vierbein: gauge choice}

After conformal transformation for the vierbein \eqref{all_25} we have  \be e^i_\mu = \delta^i_\mu + \epsilon_{\mu}^i \nonumber  \;,  \ee
where $\epsilon_{\mu}^i$ is the first-order quantity. Since the contraction of $e^i_\mu$ should give the Minkowski metric,
\be   e_{i\mu}e_j^\mu=\eta_{ij} \;,  \nonumber  \ee   one has the following expression for $e_j^\mu$,
\be   e_j^\mu=\delta_j^\mu -\epsilon_{i\nu}\delta_j^\nu\delta^{i\mu}  \;. \label{all_30} \ee
It is worth noting that the local Lorentz invariance described in Sec.~\ref{BI} can be used to make $\epsilon_{i\nu}$ symmetric: the local Lorentz transformation \eqref{all_105} adds to $\epsilon_{i\nu}$ an antisymmetric parameter $\omega_{i\nu}$. We use the gauge $$e_{\mu\nu} = e_{\nu\mu}$$ in what follows. This completely fixes the freedom under the Lorentz gauge transformations.  Then the expression \eqref{all_30} can be written as  \be e_j^{\mu} =
\delta_j^{\mu} - \epsilon_j^{\mu} \; , \;\; \text{where }\;  \epsilon_j^{\mu}
\equiv \eta^{\mu \nu} \epsilon_{j\nu} \;, \nonumber  \ee
and the metric perturbation is
\be h_{\mu \nu} =
e^i_\mu e_{i \nu} - \eta_{\mu \nu} = 2\epsilon_{\mu \nu} \nonumber \ee
Furthermore, the space-time gauge invariance can be used to choose the gauge $$e_{0 a}=0$$ and conformal Newtonian gauge. 

In this gauge, the vierbein perturbations are decomposed into helicity components as follows:
\begin{align*}
&\epsilon_{00}=-\Phi \;, \\
&\epsilon_{0a}=0  \;, \\
&\epsilon_{ab}=\Psi\delta_{ab} +i(k_{a}W_{b}+k_{b} W_{a})+\pi_{ab} \;,
\end{align*}
where $W_{a}$ is a transverse vector (vector sector), $\pi_{ab}$ is a transverse traceless tensor (tensor sector), while $\Phi$ and $\Psi$ belong to the scalar sector.

After gauge fixing, there are altogether 6 components of tensor $\epsilon_{ij}$: 2 tensor components in transverse traceless $\pi_{ab}$, 2 components in vector sector coming from the transverse vector $W_{a}$ and 2 scalar components $\Phi$ and $\Psi$.

\section{Scalar Sector} \label{SSr}
Our purpose is to study whether the theory exhibits the Boulware-Deser phenomenon and/or other instabilities in self-accelerating background. The known examples show that possible problems occur in the scalar sector of perturbations. Therefore, in this paper we study the scalar sector.

The expressions entering the linearized field equations and manipulations with these equations are cumbersome. We have designed a computer code to check the analytical part of the procedure and complete the manipulations. The code is available at \cite{code}.

\subsection{Field equations}
The explicit expressions for the quantities $F_{ijkl}$, $P_{ij}$, $S_{ijk}$, etc., that enter the linearized field equations \eqref{all_21}, \eqref{all_22} are given in Appendix B. They are used to derive the independent linearized equations in the scalar sector. It is convenient to introduce the notation
\be
\beta =\al e^{2\phi}   \; , \quad   \tilde{\beta} =\talpha e^{2\phi}   \; , \quad   {\mathfrak f} = f\e^\phi  \; , \quad   {\mathfrak g} = g\e^\phi\;,  \quad  \Lambda = \lambda\e^\phi  \; ,   \label{all_203}
\ee
where $\phi(\eta)$ is defined in \eqref{all_401}. We begin with eq.~\eqref{all_21}. It has two free indices, $i$ and $j$. The general form of the scalar part of its left hand side is
\be
{\cal G}_{ij} = \begin{cases}
{\cal G}^{(00)} \;, \quad &i=j=0 \\
ik_a{\cal G}^{(a0)} \;, \quad  &i=a, j=0 \\
ik_a{\cal G}^{(0a)} \;, \quad  &i=0, j=a \\
\delta_{ab}{\cal G}^{(\delta)} - k_ak_b{\cal G}^{(k\otimes k)} +  i\epsilon_{abc}k_c{\cal G}^{(\epsilon k)} \;, \quad &i=a, j=b 
\end{cases}
\label{all_204}
\ee
where ${\cal G}^{(00)}$, ..., ${\cal G}^{(\epsilon k)}$ are scalar functions. Their explicit forms are given in eqs.~\eqref{c1} - \eqref{c6} of Appendix C, where the relation \eqref{all_19} is used.

Equation \eqref{all_22} has three free indices, $i$, $j$, $k$, and is antisymmetric in $i$, $j$. It breaks up into the components $(0a0)$, $(ab0)$, $(0ab)$ and $(abc)$. In terms of scalar functions one has
\be
{\cal T}_{ijk} = \begin{cases}
ik_a{\cal T}^{(0a0)} \;, \quad &i=0, j=a, k=0 \\
i\epsilon_{abc}k_c{\cal T}^{(ab0)} \;, \quad  &i=a, j=b, k=0 \\
\delta_{ab}{\cal T}^{(\delta)} - k_ak_b{\cal T}^{(k\otimes k)} + i\epsilon_{abc}k_c{\cal T}^{(\epsilon k)} \;, \quad  &i=0, j=a, k=b \\
\epsilon_{abc}{\cal T}^{(\epsilon)} - \epsilon_{abd}k_ck_d{\cal T}^{(\epsilon k \otimes k)} +  i(\delta_{ac}k_b - \delta_{bc}k_a){\cal T}^{(\delta \otimes k)} \;, \; &i=a, j=b, k=c
\end{cases}
\label{all_205}
\ee
To write this decomposition, the following identity is instrumental,  \be \epsilon_{abc} = \epsilon_{abd} \frac{k_c k_d}{k^2} +
\l k_a \epsilon_{bcd} - k_b \epsilon_{acd} \r \frac{k_d}{k^2} \;.
\nonumber \ee
Thus, eq.~\eqref{all_22} leads to 8 linear equations, eqs.~\eqref{c7} - \eqref{c14}. Their explicit form is given in Appendix C.

In total we have 14 field equations: 6 of them are second order in time derivatives and 8 are first order, see Appendix C. The second order equations are eqs.~\eqref{c5}, \eqref{c9}, \eqref{c10}, \eqref{c11}, \eqref{c13} and \eqref{c14}, while the first order equations are eqs.~\eqref{c1} - \eqref{c4}, \eqref{c6} - \eqref{c8} and \eqref{c12}.

\subsection{The use of generalized Bianchi identities}
Due to the generalized Bianchi identities, not all of the field equations are independent. To see which equations may be safely ignored, let us write the generalized Bianchi identities in $(3+1)$-decomposition. The identity \eqref{all_45} gives two non-trivial identities for $\mu=0$ and for $\mu=a$, the latter being proportional to $k_a$. The identity \eqref{all_66} gives two other identities, one for $(i,j)=(0,a)$ and another for $(i,j)=(a,b)$, which are proportional to $k_a$ and $\epsilon_{abc}k_c$, respectively. 

Recall that we have made the conformal transformation \eqref{all_25}, while Bianchi identities \eqref{all_45}, \eqref{all_66} are written before the conformal transformation. In particular, $\hat{{\cal G}}_{ij}$ and $\hat{{\cal T}}_{ij}$ in \eqref{all_45}, \eqref{all_66} are the left hand sides of the general equations \eqref{all-1} and \eqref{all-2}, while we are working with equations \eqref{all_21}, \eqref{all_22} which we have derived after extracting the conformal factor. 

After conformal transformation we get the identity \eqref{all_45} with $\mu=0$:
\begin{align}
& 2{\cal G}^{(a0)}k^2 - 2\Lambda {\cal G}^{(00)} + 6\Lambda {\cal G}^{(\delta)} + 2({\cal G}^{(00)})^\prime - 2\Lambda {\cal G}^{(k\otimes k)}k^2 + 6\Lambda {\mathfrak f} {\cal T}^{(\delta)} \nonumber   \\
&- 6\Lambda {\mathfrak g} {\cal T}^{(\epsilon)} - 2 \Lambda {\mathfrak f} k^2 {\cal T}^{(k\otimes k)} + 2 \Lambda {\mathfrak g} k^2 {\cal T}^{(\epsilon k \otimes k)}  \equiv 0  \;,     \label{id1}
\end{align}
and with $\mu=a$ (which is proportional to $k_{a}$):
\begin{align}
&  -2{\cal G}^{(\delta)} + 2({\cal G}^{(0a)})^\prime + 2{\mathfrak g}{\cal T}^{(\epsilon)} -  2{\mathfrak f}{\cal T}^{(\delta)} + 2{\mathfrak f}({\cal T}^{(0a0)})^\prime - 2{\mathfrak g}({\cal T}^{(ab0)})^\prime    \nonumber \\
& + 2{\cal G}^{(k\otimes k)}k^2 + 2{\mathfrak f}k^2{\cal T}^{(k\otimes k)} - 2{\mathfrak g}k^2{\cal T}^{(\epsilon k \otimes k)} - 2\Lambda {\mathfrak g} {\cal T}^{(ab0)} + 2\Lambda {\mathfrak f} {\cal T}^{(0a0)} \equiv 0 \;.     \label{id2}
\end{align}
The identity \eqref{all_66} breaks up into 2 scalar identities:
with $i=a$, $j=0$ (which is proportional to $k_{a}$):
\be
  {\cal G}^{(0a)} -   {\cal G}^{(a0)} -  {\cal T}^{(k\otimes k)}k^2 + 2{\mathfrak f} {\cal T}^{(\delta\otimes k )} + 2{\mathfrak g}{\cal T}^{(\epsilon k)}     - ({\cal T}^{(0a0)})^\prime   +  {\cal T}^{(\delta)}  \equiv 0    \;,       \label{id3}
\ee
and with $i=a$, $j=b$ (which is proportional to $\epsilon_{abc}k_{c}$):
\be
   {\cal T}^{(\epsilon k \otimes k)}k^2 + 2{\mathfrak f}{\cal T}^{(\epsilon k)} - 2{\mathfrak g} {\cal T}^{(\delta\otimes k )}    - 2{\cal G}^{(\epsilon k)} -  {\cal T}^{(\epsilon)} + ({\cal T}^{(ab0)}  )^\prime    \equiv 0 \;.        \label{id4}
\ee
Here ${\cal G}^{(...)}$ and ${\cal T}^{(...)}$ are the left hand sides of the field equations, as defined in \eqref{all_204} and \eqref{all_205}, and their explicit forms are given in eqs.~\eqref{c1} - \eqref{c14}. We continue to use the notation \eqref{all_203}.

Thus, we have 4 scalar identities. The system \eqref{id1}-\eqref{id4} can be used to express ${\cal G}^{(\delta)}$, ${\cal T}^{(k\otimes k)}$, ${\cal T}^{(\delta\otimes k )}$ and ${\cal T}^{(\epsilon)}$  (eqs.~\eqref{c5}, \eqref{c9}, \eqref{c13} and \eqref{c14}, respectively) in terms of other components. Therefore, eqs.~\eqref{c5}, \eqref{c9}, \eqref{c13} and \eqref{c14} (which are second order) can be ignored.

\subsection{Number of propagating modes}
In Minkowski background, the scalar sector has two propagating modes. One of them is the Lorentz scalar of mass \eqref{all_90}, and another is helicity-0 part of the spin-2 excitation of mass \eqref{all_80}. The purpose of this Section is to show that the scalar sector about the self-accelerating solution \eqref{all_81} also has two modes. Thus, the Bouleware--Deser phenomenon does not occur.

In this paper we do not consider a special case $c_3 = c_4$, i.e. we impose the condition \eqref{all_340}, which implies \be 2c_3 \neq -3c_5 \label{all_402} \;. \ee We will see in due course that the case $c_3=c_4= - \frac{3}{2}c_5$ is special and requires separate treatment which we do not attempt in this paper.

Let us study the system of 10 equations: \eqref{c1} - \eqref{c4}, \eqref{c6} - \eqref{c8}, \eqref{c10} - \eqref{c12}. Two of these equations, \eqref{c10} and \eqref{c11}, are second order. However, we can replace them by first order equations. To this end, we combine ${\cal T}^{(\delta)}$, eq.~\eqref{c10} and ${\cal T}^{(\epsilon k)}$, eq.~\eqref{c11} with remaining equations and their time derivatives in the following way:
\begin{align}
{\cal D}^{(1)} \equiv &\; 3{\cal T}^{(\delta)} -\frac{1}{2\Lambda {\mathfrak f}}\left [  \Lambda {\cal G}^{(00)} - 3\Lambda ({\cal G}^{(0a)})^\prime  - 3\Lambda^2 {\mathfrak f}{\cal T}^{(0a0)} - 3\Lambda {\mathfrak f}({\cal T}^{(0a0)})^\prime    \right. \nonumber \\
&  - k^2{\cal G}^{(a0)}  - 2\Lambda k^2{\cal G}^{(k\otimes k)} - ({\cal G}^{(00)})^\prime + 3\Lambda^2 {\mathfrak g}{\cal T}^{(ab0)} \nonumber \\
& \left. + 2\Lambda {\mathfrak g} k^2{\cal T}^{(\epsilon k \otimes k)} + 3\Lambda {\mathfrak g}({\cal T}^{(ab0)})^\prime   \right ]   \label{all_209}
\end{align}
and
\begin{align}
{\cal D}^{(2)} \equiv & \; 2{\cal T}^{(\epsilon k)} -\frac{1}{\Lambda {\mathfrak f}k^2}\left [      2\Lambda {\mathfrak f}k^2{\cal T}^{(ab0)} -2 {\mathfrak f}^2\Lambda k^2{\cal T}^{(\epsilon k \otimes k)} -2 \Lambda {\mathfrak f}k^2({\cal T}^{(\epsilon k \otimes k)})^\prime  + {\mathfrak g}\Lambda{\cal G}^{(00)}    \right.  \nonumber  \\
&  - 3{\mathfrak g}\Lambda({\cal G}^{(0a)})^\prime - {\mathfrak g}({\cal G}^{(00)})^\prime - 3{\mathfrak g}\Lambda^2{\mathfrak f}{\cal T}^{(0a0)} - 3{\mathfrak g}\Lambda {\mathfrak f}({\cal T}^{(0a0)})^\prime  - {\mathfrak g}k^2{\cal G}^{(a0)}   \nonumber  \\
&\left.  - 2{\mathfrak g}\Lambda k^2{\cal G}^{(k\otimes k)}  + 3\Lambda^2 {\mathfrak g}^2{\cal T}^{(ab0)}  + 2\Lambda {\mathfrak g}^2k^2{\cal T}^{(\epsilon k \otimes k)} + 3\Lambda {\mathfrak g}^2({\cal T}^{(ab0)})^\prime   \right ] \;.  \label{all_210}
\end{align}
These are first order equations. Their explicit forms are given in Appendix D, eqs.~\eqref{d1}, \eqref{d2}. Therefore, at this stage we have 10 first order equations, \eqref{c1} - \eqref{c4}, \eqref{c6} - \eqref{c8}, \eqref{c12}, \eqref{d1}, \eqref{d2}, for 10 variables describing scalar perturbations.

The variables in the resulting system of equations are of two types. The variables of the first group, \be \sigma\;, \; \Phi\;,\; \xi\;, \; \theta \;, \label{all_300} \ee enter these equations without time derivatives. The variables of the second group, \be \Psi\;, \;\chi\;, \;\rho\;, \;Q\;, \;u\;, \; M \;,  \label{all_86} \ee enter with first time derivatives. 

Remarkably, two linear combinations of the 10 equations are actually algebraic and involve only the variables \eqref{all_86}. One of these combinations is
\begin{align}
&  {\cal A}[ 2(3c_5+2c_3)+96c_6  ] {\mathfrak g}(\Lambda-2{\mathfrak f})2{\mathfrak f}c_3  \nonumber \\
& + (-3c_5-2c_3)\Bigl \{  \left(  \frac{k^2}{3}{\cal G}^{(00)}-{\cal T}^{(ab0)}\Lambda {\mathfrak g}k^2+{\cal G}^{(0a)}\Lambda k^2-{\mathfrak g}A  \right) [4c_3({\mathfrak f}^2-{\mathfrak g}^2-\Lambda {\mathfrak f})-3\tbeta]  \nonumber \\
&  - 4\Lambda {\mathfrak f}^2c_3k^2({\cal G}^{(0a)}-{\cal G}^{(a0)})  \Bigr \} =0 \;,    \label{A1} 
\end{align}
where
$$
{\cal A} \equiv  2k^2\Lambda {\mathfrak f}{\cal T}^{(\epsilon k \otimes k)} + 3{\mathfrak g}\left( {\cal T}^{(0a0)}\Lambda {\mathfrak f}+\frac{1}{3}{\cal G}^{(00)}-{\cal T}^{(ab0)}\Lambda {\mathfrak g}+{\cal G}^{(0a)}\Lambda \right)  \;.
$$
Note that due to the condition \eqref{all_402}, the second term in eq.~\eqref{A1} does not vanish. The combination $(2c_3+3c_5)$ will appear repeatedly in formulas that follow. Thus, the case $c_3 = c_4 = - \frac{3}{2}c_5 $ is special indeed.

Another algebraic equation comes from the linear combination of eqs. \eqref{c1} - \eqref{c4}, \eqref{c6} - \eqref{c8}, \eqref{c12}, \eqref{d1}:
\begin{align}
& \left\{ \frac{4{\mathfrak f}c_3}{3\tbeta(3{\mathfrak g}^2-k^2)}[(2c_3+3c_5)(6{\mathfrak f}{\mathfrak g}^2 - \Lambda k^2) - 144{\mathfrak g}^2c_6(\Lambda-2{\mathfrak f})] \right.  \nonumber \\
& \left. - \frac{3\tbeta - 4({\mathfrak f}^2-{\mathfrak g}^2)c_3}{3\tbeta}(2c_3+3c_5) \right\} \l {\cal D} - {\cal B}\Omega \r  \nonumber \\
 & + \left\{    \Omega \left[  -4c_3k^2+\frac{12\tbeta({\mathfrak f}^2-{\mathfrak g}^2)c_3}{\beta} -\frac{(3c_5+24c_6){\mathfrak g}^2c_3}{c_6}     -  \left(\frac{6\tbeta }{\beta}+\frac{3c_5+24c_6}{4c_6}\right)\frac{12c_3{\mathfrak f}^2{\mathfrak g}^2}{(3{\mathfrak g}^2-k^2)}  \right]  \right. \nonumber \\
& \left.  + 2(3c_5 + 2c_3)[24c_5{\mathfrak f}-(96c_6 + 4c_3)(\Lambda-2{\mathfrak f})]  \right. \nonumber \\
& \left. - \frac{12{\mathfrak f}c_3[3\tbeta+6c_5({\mathfrak f}^2+{\mathfrak f}\Lambda-{\mathfrak g}^2)-4c_3({\mathfrak f}^2-{\mathfrak f}\Lambda-{\mathfrak g}^2)]}{3{\mathfrak g}^2-k^2}  \right\} {\cal E}{\mathfrak g} =0
 \;,   \label{A2}
\end{align}

where
\begin{align*}
\Omega \equiv & \frac{24c_5(3c_3+2c_5)(\Lambda+2{\mathfrak f})}{576c_6{\mathfrak g}^2-(3c_5+24c_6)^2{\mathfrak g}^2/(4c_6)+9\tbeta(\beta-\tbeta)/\beta} \;,  \\
{\cal E} \equiv & \frac{1}{3}\frac{8{\mathfrak f}c_3+12{\mathfrak f}c_5+192{\mathfrak f}c_6-4\Lambda c_3-6\Lambda c_5-96\Lambda c_6}{\tbeta (3{\mathfrak g}^2-k^2)}\left(-{\mathfrak g}{\cal T}^{(\epsilon k \otimes k)} + \frac{2k^2}{3}{\cal T}^{(0a0)}\right) \\
& + \frac{1}{3}\frac{6{\mathfrak g}^2c_3+9{\mathfrak g}^2c_5-2k^2c_3-3k^2c_5}{\tbeta (3{\mathfrak g}^2-k^2)}({\cal G}^{(a0)}-{\cal G}^{(0a)})  \;\text{,} \\
{\cal D} \equiv &  \frac{-12{\mathfrak f}^2{\mathfrak g}c_3+18{\mathfrak f}^2{\mathfrak g}c_5+12{\mathfrak f}{\mathfrak g}\Lambda c_3+18{\mathfrak f}{\mathfrak g}\Lambda c_5+12{\mathfrak g}^3c_3-18{\mathfrak g}^3c_5+9{\mathfrak g}\tbeta}{3{\mathfrak g}^2-k^2}\left({\cal T}^{(0a0)} - \frac{3}{2k^2}{\mathfrak g}{\cal T}^{(\epsilon k \otimes k)}\right) \\ 
& - \frac{24{\mathfrak g}^2c_3+36{\mathfrak g}^2c_5-8k^2c_3-12k^2c_5}{3{\mathfrak g}^2-k^2}{\cal G}^{(\epsilon k)}  \;\text{,} 
\end{align*}
\begin{align*}
{\cal B} \equiv & -\frac{1}{2 {\mathfrak f}\beta}\l  \vphantom{6{\mathfrak f}\tbeta {\cal G}^{(0a)}}  +3\beta\Lambda {\mathfrak f}{\cal T}^{(0a0)}-3\beta\Lambda {\mathfrak g}{\cal T}^{(ab0)}-2\beta {\mathfrak g}k^2{\cal T}^{(\epsilon k \otimes k)}  \right. \\
& \left. -2\beta {\mathfrak f}{\cal D}^{(1)}-6{\mathfrak f}\tbeta {\cal G}^{(0a)}   \r + \frac{3{\mathfrak g}}{c_6{\mathfrak f}}({\mathfrak f}c_5 +8{\mathfrak f}c_6 +2\Lambda c_6){\cal T}^{(ab0)}  \\
& + \frac{1}{4}\frac{6\beta {\mathfrak f}^2{\mathfrak g}k^2 c_5+48\beta {\mathfrak f}^2{\mathfrak g}k^2 c_6-12\beta {\mathfrak g}^3k^2 c_6+4\beta {\mathfrak g}k^4 c_6+48{\mathfrak f}^2{\mathfrak g}k^2\tbeta c_6}{c_6 {\mathfrak f}\beta(3{\mathfrak g}^2-k^2)}{\cal T}^{(\epsilon k \otimes k)} \\
& + \frac{1}{4}\frac{9\beta {\mathfrak f}{\mathfrak g}^2 c_5+72\beta {\mathfrak f}{\mathfrak g}^2 c_6+18\beta{\mathfrak g}^2\Lambda c_6-6\beta k^2\Lambda c_6+72{\mathfrak f}{\mathfrak g}^2 \tbeta c_6}{c_6 \beta(3{\mathfrak g}^2-k^2)}{\cal T}^{(0a0)} \;. 
\end{align*}
So, we have 2 algebraic equations \eqref{A1}, \eqref{A2} involving the variables \eqref{all_86} only, and 8 first order equations in which the variables \eqref{all_300} enter without time derivatives. Therefore, one can make use of eqs.~\eqref{A1}, \eqref{A2} to express 2 variables from the set \eqref{all_86} ($M$ and $\rho$) in terms of four remaining variables from this set ($\Psi$, $\chi$, $Q$, $u$); importantly, these expressions for $M$ and $\rho$ are algebraic, i.e. they do not involve their derivatives. 4 of these remaining 8 equations are used to express $\sigma$, $\Phi$, $\xi$, $\theta$ (the variables that enter the whole set of equations without time derivatives) in terms of $\Psi$, $\chi$, $Q$, $u$ and their first time derivatives. There remain 4 first order equations for the 4 variables ($\Psi$, $\chi$, $Q$, $u$). Thus, we have 2 degrees of freedom, the same number as in the Minkowski background. We conclude that the self-accelerating background does not exhibit the Boulware--Deser phenomenon.

In practice, it is convenient to express $\sigma$, $\xi$, $\Phi$, $\theta$ from eqs.~\eqref{c3}, \eqref{c4}, \eqref{c7} and \eqref{c8}, respectively, and substitute these into equations \be {\cal E}^{(1)} \equiv \eqref{c6}\;,  \quad {\cal E}^{(2)} \equiv \eqref{d1} \; , \label{all_302} \ee and the following linear combinations of the field equations:
\begin{align}
 {\cal E}^{(3)} \equiv & \frac{-k^2{\mathfrak f}}{\Lambda}\eqref{c2} - \frac{(5\Lambda + 2{\mathfrak f})k^2}{\Lambda}\eqref{c6} + 2{\mathfrak f}\eqref{d1} \; , \nonumber \\
 {\cal E}^{(4)} \equiv &\frac{-2(-4k^2{\mathfrak f}b - 6k^2{\mathfrak f} + 3\Lambda {\mathfrak f}^2 b -6\Lambda k^2b - 9\Lambda k^2)}{\Lambda }\eqref{c6} \nonumber \\
& - 2{\mathfrak f}(2b+3)\eqref{d1} +\frac{(2b+3)k^2}{\Lambda}\eqref{d2} \;, \label{all_303}
\end{align} where $$b \equiv \frac{c_3}{c_5}\;,$$ (c2) denotes the left hand side of eq.~\eqref{c2}, etc.
 After that we express $M$, $\rho$, $M^\prime$ and $\rho^\prime$ from the algebraic equations \eqref{A1}, \eqref{A2} and their time derivatives, and substitute these in equations ${\cal E}^{(1)}$,  ${\cal E}^{(2)}$,  ${\cal E}^{(3)}$,  ${\cal E}^{(4)}$. Thus, our linearized system reduces to the four first order equations ${\cal E}^{(1)}$,  ${\cal E}^{(2)}$,  ${\cal E}^{(3)}$,  ${\cal E}^{(4)}$ written in terms of variables $\Psi$, $\chi$, $Q$, $U$.

\subsection{Limit of small $\Lambda$}

Let us now consider the limit of small $\lambda$ and not so small $f$, eq.~\eqref{all_326}. In this case the parameters $c_5$ and $c_6$ scale with $\lambda$ as given by \eqref{all_301}.
We also assume that $|c_3| \sim |c_4| \sim |c_5|$, see \eqref{all_13}. 

The relations between ${\mathfrak g}$, $c_5$, $c_6$ and $\Lambda$, ${\mathfrak f}$, $\beta$, $\tbeta$ have the same form as in eq.~\eqref{all_201}, with the substitution $$ f \longrightarrow {\mathfrak f}\;, \quad \alpha \longrightarrow \beta \;, \quad \text{etc.}$$
In what follows we make use of these relations, and write equations in terms of (time-dependent) parameters are $\Lambda$, ${\mathfrak f}$, $\beta$, $\tbeta$. It is worth noting that we will encounter cancellations, so the leading order expressions \eqref{all_301} are not always sufficient. The higher order expressions are obtained by making use of the code \cite{code}. 

Now, consider the limit in which $\Lambda$ is the smallest parameter of the problem. For small $\Lambda$, parameters ${\mathfrak f}$, ${\mathfrak g}$, $\Lambda$, $\beta$, $\tbeta$ have slow dependence on time. We seek for solutions in the WKB form $F \sim e^{i\int\omega d\eta}$, where $F$ is any of the
variables $\Psi$, $\chi$, $Q$, $u$, and
\be
\omega \gg \Lambda, \quad k^2 \equiv k_ak_a \gg \Lambda^2 \;.  \label{all_103}
\ee
Then the time derivative is replaced by
\be
\d_0  \equiv i\omega \;.  \nonumber
\ee

In this way we obtain the system of 4 linear homogeneous algebraic equations \eqref{all_302}, \eqref{all_303} for $\Psi$, $\chi$, $Q$, $u$, which determine the dispersion relations $\omega_i=\omega_i(k), i=1,\dots,4$. Equating the determinant of the system to zero we get a fourth order equation for $\omega$. To see that there are exponentially growing modes, we consider two ranges of momenta,
\be  k^2 \ll \Lambda {\mathfrak f} \label{all_329} \ee and \be {\mathfrak f}^3\Lambda \ll k^4 \ll \frac{{\mathfrak f}^5}{\Lambda} \;. \label{all_321}  \ee
Note that $k$ is conformal momentum, while $\Lambda$ and ${\mathfrak f}$ are conformally related to the physical parameters of the solution. Equation \eqref{all_329}, written in physical terms, is $$ \left( \frac{k}{a(\eta)}\right) ^2 \ll \lambda f \;,$$ and similarly for \eqref{all_321}.

We are going to see that in the range \eqref{all_329} there are instabilities $$\omega \sim -i\frac{{\mathfrak f}^2}{\Lambda} \quad \text{and} \quad \omega \sim -i\sqrt{\frac{\Lambda^3{\mathfrak f}}{k^2}} \;,$$ while in range \eqref{all_321} there are instabilities $$\omega \sim -i\frac{{\mathfrak f}^2}{\Lambda} \;.$$

We first consider the case \eqref{all_321}. The fourth order equation for $\omega$ (obtained by equating the determinant to zero), to the  leading order in $\Lambda$, is:
\begin{align}
& -(108{\mathfrak f}^4b+72{\mathfrak f}^4b^2+45{\mathfrak f}^2k^2+18{\mathfrak f}^2b^2k^2+51{\mathfrak f}^2bk^2-2bk^4-3k^4)(6{\mathfrak f}^2+k^2)\omega^4\Lambda^2 \nonumber \\
&   -i(-3k^4+27{\mathfrak f}^2k^2-2bk^4+33{\mathfrak f}^2bk^2+36{\mathfrak f}^4b+14{\mathfrak f}^2b^2k^2+24{\mathfrak f}^4b^2)(6{\mathfrak f}^2+k^2){\mathfrak f}^2\omega^3\Lambda  \nonumber \\
& -2{\mathfrak f}^6(6{\mathfrak f}^2+k^2)(2b+3)(12{\mathfrak f}^2b+bk^2+3k^2) \omega^2 + 8i(12{\mathfrak f}^2b+bk^2+3k^2)(2b+3)k^2{\mathfrak f}^6\omega\Lambda \nonumber \\
&  - 8{\mathfrak f}^6(18{\mathfrak f}^2-k^2)(2b+3)(12{\mathfrak f}^2b+bk^2+3k^2)\Lambda^2 =0 \;. \label{DetEq}
\end{align}
This expression is obtained by making use of the code \cite{code} and accounts for cancellations between dominant terms in $\Lambda$. Equation \eqref{DetEq} is fourth order in $\omega$, so there are formally 4 roots. Some of these roots may not obey $\omega \gg \Lambda$, which is our original assumption, see \eqref{all_103}. Our approach is to find all 4 roots of eq.~\eqref{DetEq} and then figure out the roots obeying $\omega \gg \Lambda$. In this way we gain confidence that no relevant solutions to eq.~\eqref{DetEq} are lost. To this end, we consider limiting cases.
  
\begin{itemize}
\item $\omega \ll \Lambda$. Then all the terms containing $\omega$ are small and eq.~\eqref{DetEq} is simply:
\begin{equation}
- 8{\mathfrak f}^6(18{\mathfrak f}^2-k^2)(2b+3)(12{\mathfrak f}^2b+bk^2+3k^2)\Lambda^2 =0 \label{D1}
\end{equation}
Equation \eqref{D1} does not involve $\omega$, so there are no roots.
\item $\omega \sim \Lambda$. In this case eq.~\eqref{DetEq} becomes quadratic,
\begin{align}
& -2{\mathfrak f}^6(6{\mathfrak f}^2+k^2)(2b+3)(12{\mathfrak f}^2b+bk^2+3k^2) \omega^2 \nonumber \\
&+ 8i(12{\mathfrak f}^2b+bk^2+3k^2)(2b+3)k^2{\mathfrak f}^6\omega\Lambda \nonumber \\
&  - 8{\mathfrak f}^6(18{\mathfrak f}^2-k^2)(2b+3)(12{\mathfrak f}^2b+bk^2+3k^2)\Lambda^2 =0 \;, \nonumber
\end{align}
and has two roots: 
\be
 \omega_{1,2} = 2i\frac{k^2 \pm 2\sqrt{27{\mathfrak f}^4+3{\mathfrak f}^2k^2}}{6{\mathfrak f}^2+k^2}\Lambda \;. \nonumber
\ee
These roots, however, do not obey $\omega \gg \Lambda$ and thus are irrelevant.
\item $\Lambda \ll \omega \ll \frac{{\mathfrak f}^2}{\Lambda}$.
Then $\omega\Lambda \ll {\mathfrak f}^2$ and 
\begin{equation*}
 \omega^3\Lambda  \ll \omega^2{\mathfrak f}^2 \;,\;  \omega^4\Lambda^2  \ll \omega^2{\mathfrak f}^4 \;.
\end{equation*}
In eq.~\eqref{DetEq}, only the third term survives, and we have simply
\be  -2{\mathfrak f}^6(6{\mathfrak f}^2+k^2)(2b+3)(12{\mathfrak f}^2b+bk^2+3k^2) \omega^2=0\; . \nonumber \ee There are no roots.
\item $\omega \sim \frac{{\mathfrak f}^2}{\Lambda}$. In this case terms proportional to $\omega\Lambda$ and $\Lambda^2$ are much smaller than others, and eq.~\eqref{DetEq} has the following form:
\begin{align}
& -(108{\mathfrak f}^4b+72{\mathfrak f}^4b^2+45{\mathfrak f}^2k^2+18{\mathfrak f}^2b^2k^2+51{\mathfrak f}^2bk^2-2bk^4-3k^4)(6{\mathfrak f}^2+k^2)\omega^4\Lambda^2 \nonumber \\
&   -i(-3k^4+27{\mathfrak f}^2k^2-2bk^4+33{\mathfrak f}^2bk^2+36{\mathfrak f}^4b+14{\mathfrak f}^2b^2k^2+24{\mathfrak f}^4b^2)(6{\mathfrak f}^2+k^2){\mathfrak f}^2\omega^3\Lambda  \nonumber \\
& -2{\mathfrak f}^6(6{\mathfrak f}^2+k^2)(2b+3)(12{\mathfrak f}^2b+bk^2+3k^2) \omega^2  =0 \;. \label{D3}
\end{align}
Equation \eqref{D3} has two roots:
\begin{align}
& \omega_{3} = -i\frac{{\mathfrak f}^2}{\Lambda}  \;, \label{all_403} \\
& \omega_{4} = 2i\frac{(12{\mathfrak f}^2b+bk^2+3k^2)(2b+3){\mathfrak f}^4}{\Lambda(108{\mathfrak f}^4b+72{\mathfrak f}^4b^2+45{\mathfrak f}^2k^2+18{\mathfrak f}^2b^2k^2+51{\mathfrak f}^2bk^2-2bk^4-3k^4) } \;. \label{all_404}
\end{align}
\item $\omega \gg \frac{f^2}{\Lambda}$. In this case terms with $\omega$ are much larger than others. Eq.~\eqref{DetEq} is:
\begin{equation*}
-(108{\mathfrak f}^4b+72{\mathfrak f}^4b^2+45{\mathfrak f}^2k^2+18{\mathfrak f}^2b^2k^2+51{\mathfrak f}^2bk^2-2bk^4-3k^4)(6{\mathfrak f}^2+k^2)\omega^4\Lambda^2 = 0 \;.
\end{equation*}
It has no roots.
\end{itemize}
We see that "frequency" \eqref{all_403}, and also \eqref{all_404} in a certain range of $k^2$, are of order $$\omega_{3, 4} \sim -i\frac{{\mathfrak f}^2}{\Lambda} \;.$$ They are pure imaginary and correspond to instabilities.

Now let us consider the case $k^2 \ll \Lambda {\mathfrak f}$, eq.~\eqref{all_329}. To the leading order in $\Lambda$, the fourth order equation for the determinant is as follows,
\begin{align}
& 6b\beta k^2\omega^4\Lambda^2+(5 b\beta k^4+3 \beta k^4)\Lambda^2\omega^3+24b\beta k^2\omega^2{\mathfrak f}^4 \nonumber \\
& + 8 \Lambda {\mathfrak f}^3 \tbeta k^4 \omega +864b\tbeta\Lambda^3{\mathfrak f}^5 =0 \;. \label{all_304}
\end{align}
Let us again consider limiting cases.
\begin{itemize}
\item $\omega \lesssim \Lambda$. In this case the term of zeroth order in $\omega$ is the largest, and eq.~\eqref{all_304} is simply $$ 864b\tbeta\Lambda^3{\mathfrak f}^5 =0   \; .$$ So, there are no roots.
\item $\Lambda \ll \omega \ll \frac{{\mathfrak f}^2}{\Lambda}$. In this case terms with $\omega^2$ and $\omega^0$ are much larger than others, and eq.~\eqref{all_304} is:
$$ 24b\beta k^2\omega^2{\mathfrak f}^4 + 864b\tbeta\Lambda^3{\mathfrak f}^5 =0 \;.$$
There are two roots,
\be \omega_{5} = - 6\sqrt{\frac{\tbeta {\mathfrak f \Lambda}}{\beta k^2}}i\Lambda \;, \quad \omega_{6} = 6\sqrt{\frac{\tbeta {\mathfrak f \Lambda}}{\beta k^2}}i\Lambda \;.  \label{all_320} \ee
\item $\omega \sim \frac{{\mathfrak f}^2}{\Lambda}$. Then only the first and third terms in eq.~\eqref{all_304} survive, and we have
$$ 6b\beta k^2\omega^4\Lambda^2 +24b\beta k^2\omega^2{\mathfrak f}^4 =0  \;. $$ This equation has 2 roots: $$ \omega_{7} = -\frac{2i{\mathfrak f}^2}{\Lambda} \;, \quad \omega_{8} = \frac{2i{\mathfrak f}^2}{\Lambda} \;. $$
\item $\omega \gg \frac{{\mathfrak f}^2}{\Lambda}$. Then the first term in eq.~\eqref{all_304} is the largest, and we have
$$ 6b\beta k^2\omega^4\Lambda^2=0 \;. $$ There are no roots.
\end{itemize}
The frequencies $\omega_{5}$ and $\omega_{7}$ correspond to the instabilities. This completes our analysis of momenta \eqref{all_329}, \eqref{all_321}.

The instabilities $\omega_3$, $\omega_4$ and $\omega_7$ do not necessarily kill the model. They can be sent beyond a UV cutoff necessarily present in the model by imposing the condition
\be  \omega \sim \frac{{\mathfrak f^2}}{\Lambda} \gtrsim M_{UV}  \;,  \nonumber \ee
where $M_{UV}$ is the cutoff scale. 

On the other hand, one cannot get rid of the instability $\omega_5$. Indeed, we study background \eqref{all_141}, thus we have to assume that ${\mathfrak f} \ll M_{UV}$. This means, taking into account the second inequality of \eqref{all_103}, that
\be  \omega = \frac{\Lambda}{k}\sqrt{\Lambda{\mathfrak f}} \ll M_{UV} \;.  \nonumber \ee
Thus, we cannot push it beyond the UV-cutoff. Furthermore, for low enough $k$ (but still $k \gg \Lambda$), the time scale of instability is short. Indeed, the case we consider in this paper is
\be
f \gg \lambda \log^2A \;, \label{all_405}
\ee
where $A$ is the initial amplitude of perturbation. In this case, the exponential growth of perturbation makes it large (and non-linear) in the Hubble time. 

The analysis given in this Section shows that the self-accelerating solution \eqref{all_141}, under assumption \eqref{all_405}, is unstable because of the exponentially growing mode \eqref{all_320}.

\section{Discussion}
Let us summarize the results. We studied the stability of the self-accelerating solution \eqref{all_81} of the model \eqref{all_2} at the linearized level under assumptions \eqref{all_340} - \eqref{all_13}. We made \mbox{$(3+1)$}-decomposition of small perturbations and considered the scalar sector. We found that the number of degrees of freedom in the scalar sector is the same as in Minkowski background, i.e. there are no Bouleware--Deser modes. We further studied the spectrum of perturbations under assumptions \eqref{all_328} and \eqref{all_326} and found the exponentially growing modes \be  \omega_{3, 4, 7} \sim -i\frac{{\mathfrak f}^2}{\Lambda} \qquad \text{and} \qquad \omega_5 \sim -i\sqrt{\frac{\Lambda^3{\mathfrak f}}{k^2}}  \;. \label{all_406} \ee The first instability can be pushed beyond the UV cutoff and thus is not fatal, whereas the second one cannot be sent beyond the UV cutoff and makes this solution unstable. There remains a possibility that the self-accelerating solution is healthy for small $f \sim \lambda$. Indeed, taken at face value, the results \eqref{all_406} suggest that in that case all instabilities may be slow, $$  \omega_5 \sim \frac{\Lambda^2}{k} \ll \Lambda \;, \qquad \omega_{3,4,7} \sim \frac{{\mathfrak f}^2}{\Lambda} \sim \Lambda  \;. $$ We emphasize, however, that our analysis is not valid for $f \sim \lambda$, so this case needs a separate study. Note that in this case one has $g \sim \lambda$, according to the third of eq.~\eqref{all_201}, so the entire background torsion is small. The small--$f$ regime can be achieved provided that couplings $c_5$, $c_6$ are of order $$ c_5 \sim c_6 \sim O(\lambda^{-2}) \;. $$ We plan to turn to this case in future.

Finally, in this paper we considered the case $c_3 \neq c_4$, see \eqref{all_340}. As can be seen from eq.~\eqref{A1} and subsequent equations, our analysis does not apply if the condition \eqref{all_340} is violated, i.e.
$$  c_3=c_4=-\frac{3}{2}c_5 \; \qquad \text{or, in other words, }\qquad b = -\frac{3}{2} \;. $$
In that case the dynamics of small perturbations can be significantly different.

\section*{Acknowledgements}
The author is deeply indebted to Valery Rubakov for numerous fruitful and productive discussions. Special thanks are to Fyodor Tkachev for his advice in analytical computing, and to Leonid Ledentsov for a few key comments. The author thanks Konstantin Astapov, Emin Nugaev, Victoria Volkova and Inar Timiryasov for numerous suggestions and comments. This work has been supported by Russian Science Foundation grant 14-22-00161.

\newpage

\section*{Appendix A}

The object $H_{ijk}$ emerges as follows. When varying the action
with respect to $A_{ij\mu}$ one makes use of the expression \be
\delta F_{ijkl} = e^\mu_k e^\nu_l \l {\cal D}_\mu \delta A_{ij\nu} -
{\cal D}_\nu \delta A_{ij\mu}\r \;, \nonumber \ee where one can
understand ${\cal D}_\mu$ as ``total'' covariant derivative, \be
  {\cal D}_\mu \delta A_{ij\nu} =  \d_\mu \delta A_{ij\nu}
+ A_{ik\mu} \delta A_{kj\nu} +
 A_{jk\mu} \delta A_{ik\nu} - \Gamma_{\mu \nu}^\lambda
A_{ij\lambda} \;, \nonumber \ee and $\Gamma_{\mu \lambda}^\nu$ is Riemannian
connection (Christoffel symbol). One integrates the variation of the action by parts and encounters the
expression \be {\cal D}_\mu \l  e^\mu_k e^\nu_l\r - {\cal D}_\mu \l  e^\nu_k
e^\mu_l\r \;. \nonumber \ee One recalls that \be \d_\mu e^\nu_i + \Gamma^{\nu}_{\mu \Lambda} e^\lambda_i +
\omega_{ij\mu} e^\nu_j = 0 \;, \nonumber \ee where $\omega_{ij\mu}$ is
Riemannian spin connection. Now, ${\cal D}_\mu$ involves \be A_{ij\mu} =
\omega_{ij\mu} + K_{ij\mu} \; , \nonumber \ee and therefore \be e_{k \nu }
\left[{\cal D}_\mu \l  e^\mu_i e^\nu_j\r - {\cal D}_\mu \l  e^\nu_i e^\mu_j\r
\right] = -\frac{2}{3\talpha} H_{ijk} \nonumber \ee with the
definition of $H_{ijk}$ given in \eqref{all_41}. So, one has \be \delta \l \int~d^4x~e F
\r = \int~d^4x~e~ \frac{2}{3\talpha}  H_{ijk} e^{k \mu} \delta
A_{ij\mu} \; , \nonumber \ee
\begin{align*}
\delta \l \int~d^4x~ e F_{ij}F_{ij} \r=  \int~d^4x~e~ & \left[-\l D_i
F_{jk} - D_j F_{ik} \r + \l \eta_{ik} D_l F_{jl} - \eta_{jk} D_j
F_{il} \r \right.
\\
& \left. + \frac{2}{3\talpha} \l H_{ilk} F_{jl} - H_{jlk}F_{il} \r
\right]   e^{k \mu} \delta A_{ij\mu} \;,
\end{align*}
etc. In this way one obtains Eq.~\eqref{all-2}.

\newpage

\section*{Appendix B}

Linearized Riemannian connection for symmetric $\epsilon_{ij}$
reads (after conformal transformation):
\be
\omega_{ijk} = \d_j \epsilon_{ik} - \d_i \epsilon_{jk} \; \nonumber
\ee
and therefore we have in the scalar sector
\be
\omega_{0a0} = - i k_a \Phi \; , \quad 
\omega_{0ab}  = -  \Psi^\prime \delta_{ab} \; , \quad 
\omega_{ab0} = 0 \; , \quad 
\omega_{abc}  = -i (k_a \delta_{bc} - k_b \delta_{ac}) \Psi \;.  \nonumber
\ee
The linearized Ricci tensor reads
\be
R_{00}  = -3 \Psi^{\prime \prime} - k^2 \Phi \; , \quad 
R_{0a} = -2 i k_a \Psi^\prime \; , \quad 
R_{ab} = (\Psi^{\prime \prime} + k^2 \Psi) \delta_{ab} + k_a k_b 
(\Psi + \Phi) \;, \nonumber
\ee
and the Ricci scalar is \be R = 6 \Psi^{\prime \prime} + 4 k^2 \Psi +
2 k^2 \Phi  \;. \nonumber \ee The additional term in eq. \eqref{all_21} that emerges due to
conformal transformation, to the linear order in perturbations, breaks up into three components, 
\begin{align*}
\Delta_{00}^{(G)} &=  - \frac{3\alpha}{2} \e^{-2\phi} \l -6
\phi^{\prime \, 2} \Phi + 6 \phi^\prime \Psi^\prime \r \;,
\\
\Delta_{0a}^{(G)} = \Delta_{a0}^{(G)} &=  - \frac{3\alpha}{2}
\e^{-2\phi} \l 2i k_a \phi^\prime \Phi \r \;,
\\
\Delta_{ab}^{(G)} &=  - \frac{3\alpha}{2} \e^{-2\phi} \l - 4
\phi^\prime \Psi^\prime + 4 \phi^{\prime \prime} \Phi + 2
\phi^{\prime \, 2} \Phi \r \delta_{ab} \;.
\end{align*}
Similarly, we have additional terms in eq.~\eqref{all_22}:
\begin{align*}
\Delta^{(T)}_{0a0} = \Delta^{(T)}_{ab0} & = \Delta^{(T)}_{abc}  =
0 \;,
\\
\Delta^{(T)}_{0ab} &= \e^{-\phi} 3 \talpha \phi^\prime \Phi
\delta_{ab} \;.
\end{align*}

Since the field equations involve terms quadratic in $F_{ijkl}$, we keep both background parts and terms linear in perturbations. We proceed with $F_{ijkl}$. The explicit calculation gives:
\begin{align*}
F_{0a0b}= &\delta_{ab}[ e^{2\phi}\lambda f - \Psi^{\prime\prime} -
e^{2\phi}\lambda f (\Psi + \Phi) + \sigma^\prime ] + k_ak_b [
-\Phi
+ \chi^\prime - i\xi ]  
\\
 &+\epsilon_{abc}k_c[ -ie^{\phi}g\Phi + \rho^\prime +
e^\phi( g\xi + f\theta) ]  \;, 
\end{align*}
\begin{align*}
 F_{0abc}= &\epsilon_{abc}[ -2fge^{2\phi} + 4e^{2\phi}fg\Psi +
2e^\phi g\Psi^\prime - 2g\sigma e^\phi ] 
\\ 
 & + e^\phi k_a\epsilon_{bcd}k_d[-2fu] 
\\
 &+(k_b\delta_{ac} - k_c\delta_{ab})[ -i\Psi^\prime -
ie^\phi f\Psi + i\sigma + e^\phi fM - g\rho e^\phi ] 
\\
 &+(\epsilon_{abd}k_ck_d - \epsilon_{acd}k_bk_d)[ -i\rho
+ e^\phi(-fQ - fu - g\chi) ]  \;, 
\end{align*}
\begin{align*}
F_{aboc}= &\epsilon_{abc}\lambda ge^{2\phi}(1-\Phi-\Psi) + \epsilon_{abd}k_ck_d(Q^\prime - i\theta) \\
& + (k_a\delta_{bc} - k_b\delta_{ac})(-i\Psi^\prime - ie^\phi f \Phi +
M^\prime + e^\phi(f\xi-g\theta))
\\
 &+(k_a\epsilon_{bcd} - k_b\epsilon_{acd})k_du^\prime  \;, 
\end{align*}
\begin{align*}
F_{abcd}= &(\delta_{ac}\delta_{bd} - \delta_{ad}\delta_{bc})[
e^{2\phi}(f^2-g^2)  \\
&- 2\Psi e^{2\phi}(f^2-g^2) - 2e^\phi f \Psi^\prime +
e^\phi (2f\sigma - 2k^2 g u)]
\\
 &+ (k_ck_a\delta_{bd} - k_ck_b\delta_{ad} -
k_dk_a\delta_{bc} + k_dk_b\delta_{ac})[ \Psi + iM +
e^\phi(f\chi-gQ+gu) ] 
\\
&+ (\epsilon_{acd}k_b - \epsilon_{bcd}k_a)[ -ie^\phi g
\Psi + (f\rho + gM)e^\phi ] - ik^2u (\epsilon_{abc}k_d -
\epsilon_{abd}k_c)  \;.
\end{align*}
The linearized $F_{ij}$ reads:
\begin{align*}
F_{00} = &3[ e^{2\phi}\lambda f - \Psi^{\prime\prime} -
e^{2\phi}\lambda f (\Psi + \Phi) + \sigma^\prime ] + k^2(-\Phi +
\chi^\prime - i\xi) \;, 
\\
F_{a0} = &2k_a(-i\Psi^\prime - ie^\phi f \Phi + M^\prime +
e^\phi(\xi - g\theta))  \;, 
\\
F_{0a} = &2k_a(-i\Psi^\prime - ie^\phi f \Psi + i\sigma + e^\phi f
M - g\rho e^\phi)  \;, 
\\
F_{ab} =& \delta_{ab} [ -e^{2\phi}\lambda f + \Psi^{\prime\prime}
+ e^{2\phi}\lambda f (\Psi+\Phi) - \sigma^\prime +
2e^{2\phi}(f^2-g^2) -4\Psi e^{2\phi}(f^2-g^2) 
\\
&- 4e^\phi f \Psi^\prime + 2e^\phi(2f\sigma - 2k^2gu) +
k^2\Psi + ik^2M + e^\phi
k^2(f\chi - gQ +gu) ]
\\
&+ k_ak_b[ \Phi - \chi^\prime + i\xi + \Psi + iM +
e^\phi(f\chi - gQ + gu) ] 
\\
 &- \epsilon_{abc}k_c[ -ie^\phi g \Phi + \rho^\prime +
e^\phi(g\xi + f\theta) + ie^\phi g \Psi  - (f\rho + gM)e^\phi -
ik^2u] \;.
\end{align*}
This expressions determine also the tensor $P_{ij} = c_3F_{ij} + c_4F_{ji}$.
The scalars $F$ and $\epsilon\cdot F$ are
\begin{align*}
F = & \; -6[ e^{2\phi}\lambda f - \Psi^{\prime\prime} -
e^{2\phi}\lambda f
(\Psi+\Phi) + \sigma^\prime ] - 2k^2(-\Phi + \chi^\prime - i\xi) \\
&+ 3[ 2e^{2\phi}(f^2-g^2) - 4\Psi e^{2\phi}(f^2-g^2) -
4e^\phi f \Psi^\prime + 2e^\phi(2f\sigma - 2k^2gu) ] \\
 &+ 4[ k^2\Psi + ik^2M + e^\phi k^2 (f\chi - gQ + gu) ] \;, \\
\epsilon\cdot F = & \;12[ -2fge^{2\phi} + 4e^{2\phi}fg\Psi + 2e^\phi
g\Psi^\prime - 2g\sigma e^\phi + \lambda g
e^{2\phi}(1-\Phi-\Psi) ] \\
 &+ 4k^2[
2u^\prime+Q^\prime-i\theta-2i\rho+2e^\phi(-fQ-fu-g\chi) ] \;.
\end{align*}
The scalar $P$ is simply $P=-3c_5F$.

The linearized covariant derivatives $D_iP_{ik}$ read
\begin{align*}
D_0P_{00} = &(c_3+c_4)[ 6\lambda^2 f e^{3\phi}(1-\Psi-2\Phi) -
3\Psi^{\prime\prime\prime} - 3e^{2\phi}\lambda f
(\Psi^\prime+\Phi^\prime) \\
 &+ 3\sigma^{\prime\prime} - k^2\Phi^\prime +
k^2\chi^{\prime\prime} - k^2i\xi^\prime ] \;, 
\\
D_aP_{00} = &ik_a(c_3+c_4)\{ 3e^{2\phi}\lambda f (-\Psi-\Phi) -
3\Psi^{\prime\prime} + 3\sigma^\prime + k^2\chi^\prime \\
 &- k^2\Phi - k^2i\xi - 3\Psi e^{2\phi}f\lambda -
4\Psi^\prime e^\phi f + 2e^\phi f [-e^\phi f \Psi + \sigma \\
 &- ie^\phi(fM-g\rho+f\xi-g\theta)-e^\phi f \Phi -
iM^\prime] \} \;,
\end{align*}

\begin{align*}
D_0P_{0a} = &k_a(c_3+c_4)[ -2\Psi^{\prime\prime} i -
2e^{2\phi}(i\Phi-\xi)(\lambda f + f^2 - g^2) ] \\
\nonumber &+ 2k_ac_3[ -if\Psi^\prime e^\phi - if\Psi\lambda
e^{2\phi} + i\sigma^\prime + \lambda e^{2\phi}(fM-g\rho) +
e^\phi(fM^\prime - g\rho^\prime) ] \\
\nonumber &+ 2k_ac_4[ -ie^\phi f\Phi^\prime - if\Phi\lambda
e^{2\phi} + M^{\prime\prime} + \lambda e^{2\phi}(f\xi-g\theta) +
e^\phi(f\xi^\prime-g\theta^\prime) ] \;,
\\
D_0P_{a0} = &k_a(c_3+c_4)[ -2\Psi^{\prime\prime} i -
2e^{2\phi}(i\Phi-\xi)(\lambda f + f^2 - g^2) ] \\
\nonumber &+ 2k_ac_4[ -if\Psi^\prime e^\phi - if\Psi\lambda
e^{2\phi} + i\sigma^\prime + \lambda e^{2\phi}(fM-g\rho) +
e^\phi(fM^\prime - g\rho^\prime) ] \\
\nonumber &+ 2k_ac_3[ -ie^\phi f\Phi^\prime - if\Phi\lambda
e^{2\phi} + M^{\prime\prime} + \lambda e^{2\phi}(f\xi-g\theta) +
e^\phi(f\xi^\prime-g\theta^\prime) ] \;,
\end{align*}

\begin{align*}
D_0P_{ab} = &\delta_{ab}(c_3+c_4)[ -\lambda^2 f 2 e^{3\phi} +
\Psi^{\prime\prime\prime} + 2\lambda^2 e^{3\phi}f(\Psi+\Phi) +
e^{2\phi}\lambda f (\Psi^\prime+ \Phi^\prime) \\
\nonumber &- \sigma^{\prime\prime} + 4\lambda e^{3\phi}(f^2-g^2) -
4\Psi^\prime e^{2\phi}(f^2-g^2) - 8\Psi(f^2-g^2)\lambda e^{3\phi}
\\
\nonumber
&- 2\lambda e^{3\phi}\Phi (2f^2-2g^2-\lambda f) - 4
e^\phi f
\Psi^{\prime\prime} - 4f\Psi^\prime\lambda e^{2\phi} \\
\nonumber &+ 2e^\phi(2f\sigma^\prime-2k^2gu^\prime) + 2\lambda
e^{2\phi}(2f\sigma-2k^2gu) + k^2\Psi^\prime + ik^2M^\prime \\
\nonumber
&+ \lambda e^{2\phi}k^2(f\chi-gQ+gu) + e^\phi
k^2(f\chi^\prime-gQ^\prime+gu^\prime) ] \\
\nonumber &+ k_ak_b(c_3+c_4)[ \Phi^\prime - \chi^{\prime\prime} +
i\xi^\prime + \Psi^\prime + iM^\prime + \lambda
e^{2\phi}(f\chi-gQ+gu) 
\\
\nonumber &+ e^\phi(f\chi^\prime-gQ^\prime+gu^\prime)
]+\epsilon_{abc}k_c(c_4-c_3)[ -ie^\phi g \Phi^\prime -
ig\Phi\lambda e^{2\phi} 
\\
\nonumber &+ \rho^{\prime\prime} +
e^\phi(g\xi^\prime+f\theta^\prime) + \lambda
e^{2\phi}(g\xi+f\theta)+ ie^\phi g\Psi^\prime + i\lambda e^{2\phi}g\Psi \nonumber \\
& - (f\rho^\prime+gM^\prime)e^\phi - (f\rho+gM)\lambda e^{2\phi} -
ik^2u^\prime ] \;,
\end{align*}

\begin{align*}
D_bP_{a0} = &k_ak_b\left\{ 2(c_3+c_4)\Psi^\prime + 2ic_4[-ie^\phi f
\Psi
+ i\sigma + e^\phi(fM-g\rho)] \right.\\
\nonumber &+ 2ic_3[-ie^\phi f \Phi + M^\prime +
e^\phi(f\xi-g\theta)] + 3\chi(c_3+c_4)e^{2\phi}\lambda f \\
\nonumber &+ e^\phi f (c_3+c_4)[\Phi-\chi^\prime+i\xi+\Psi + iM +
e^\phi(f\chi-gQ+gu)]  \\
\nonumber & \left.+ \chi(c_3+c_4)e^{2\phi}(2f^2-2g^2-\lambda f)
\right\}+ \delta_{ab}(c_3+c_4)\left\{ e^\phi f (3e^{2\phi}\lambda f -
3\Psi^{\prime\prime} \right.  \\
\nonumber &- 3e^{2\phi}\lambda f (\Psi + \Phi) +
3\sigma^\prime + k^2\chi^\prime - k^2\Phi - i\xi k^2) \\
\nonumber
&+ 3e^{2\phi}\lambda f (-\Psi^\prime + \sigma -
fe^\phi\Psi)+ e^\phi f [ -e^{2\phi}\lambda f + \Psi^{\prime\prime}
+ e^{2\phi}\lambda f (\Psi+\Phi)   \\
\nonumber & - \sigma^\prime +
2e^{2\phi}(f^2-g^2)- 4\Psi e^{2\phi}(f^2-g^2) - 4e^\phi f
\Psi^\prime + 2e^\phi(2f\sigma - 2k^2gu)   \\
\nonumber
&+ k^2\Psi + ik^2M+ e^\phi k^2 (f\chi-gQ+gu) ] \\
&\left. + (-\Psi^\prime + \sigma -
f e^\phi\Psi)e^{2\phi}(2f^2-2g^2-\lambda f)-\Psi[f^23e^{3\phi}\lambda + e^{3\phi}f(2f^2-2g^2-\lambda f)] \right\}\\
\nonumber &+ \epsilon_{abc}k_c \left\{ 3\rho(c_3+c_4)e^{2\phi}\lambda f
+ ge^\phi2(c_3+c_4)i\Psi^\prime \right. \\
& - ge^\phi2c_4[-ie^\phi f \Psi +
i\sigma + e^\phi(fM-g\rho)] - ge^\phi 2c_3[-ie^\phi f \Phi + M^\prime +
e^\phi(f\xi-g\theta)]   \\
\nonumber &+ (c_4-c_3)[-ie^\phi g \Phi + \rho^\prime +
e^\phi(g\xi+f\theta) + ie^\phi g \Psi - (f\rho+gM)e^\phi - ik^2u]e^\phi f  \\
\nonumber
&\left.+ \rho(c_3+c_4)e^{2\phi}(2f^2-2g^2-\lambda f) \right\} \;,
\end{align*}

\begin{align*}
D_bP_{0a} = &k_ak_b\{ 2(c_3+c_4)\Psi^\prime + 2ic_3[-ie^\phi f
\Psi
+ i\sigma + e^\phi(fM-g\rho)] \\
 &+ 2ic_4[-ie^\phi f \Phi + M^\prime +
e^\phi(f\xi-g\theta)] + 3\chi(c_3+c_4)e^{2\phi}\lambda f  \\
 &+ e^\phi f (c_3+c_4)[\Phi-\chi^\prime+i\xi+\Psi + iM +
e^\phi(f\chi-gQ+gu)] \\
& + \chi(c_3+c_4)e^{2\phi}(2f^2-2g^2-\lambda f)
\} \\
&+ \delta_{ab}(c_3+c_4)\{ e^\phi f [3e^{2\phi}\lambda f -
3\Psi^{\prime\prime} - 3e^{2\phi}\lambda f (\Psi + \Phi)  \\
& + 3\sigma^\prime + k^2\chi^\prime - k^2\Phi - i\xi k^2] 
+ 3e^{2\phi}\lambda f [-\Psi^\prime + \sigma -
fe^\phi\Psi]  \\
& + e^\phi f [ -e^{2\phi}\lambda f + \Psi^{\prime\prime}
+ e^{2\phi}\lambda f (\Psi+\Phi) - \sigma^\prime +
2e^{2\phi}(f^2-g^2)  \\
&- 4\Psi e^{2\phi}(f^2-g^2) - 4e^\phi f
\Psi^\prime + 2e^\phi(2f\sigma - 2k^2gu) \\
& + k^2\Psi + ik^2M  
+ e^\phi k^2 (f\chi-gQ+gu) ] \\
&+ (-\Psi^\prime + \sigma -
f e^\phi\Psi)e^{2\phi}(2f^2-2g^2-\lambda f)  \\
&-\Psi[f^23e^{3\phi}\lambda + e^{3\phi}f(2f^2-2g^2-\lambda f) ] \} \\
 &+ \epsilon_{abc}k_c \{ 3\rho(c_3+c_4)e^{2\phi}\lambda f
+ ge^\phi2(c_3+c_4)i\Psi^\prime \\
& - ge^\phi2c_3[-ie^\phi f \Psi +
i\sigma + e^\phi(fM-g\rho)] \\
 &- ge^\phi 2c_4[-ie^\phi f \Phi + M^\prime +
e^\phi(f\xi-g\theta)] \\
&+ (c_3-c_4)[-ie^\phi g \Phi + \rho^\prime +
e^\phi(g\xi+f\theta) \\
 &+ ie^\phi g \Psi - (f\rho+gM)e^\phi - ik^2u]e^\phi f\\
 & + \rho(c_3+c_4)e^{2\phi}(2f^2-2g^2-\lambda f) \} \;,
\end{align*}

\begin{align*}
D_cP_{ab} = &ik_c\delta_{ab}(c_3+c_4)[ \Psi^{\prime\prime} +
e^{2\phi}\lambda f (\Psi+\Phi) - \sigma^\prime \\
\nonumber &- 4\Psi e^{2\Phi}(f^2-g^2) - 4e^\phi f \Psi^\prime +
2e^\phi(2f\sigma - 2k^2gu) + k^2\Psi \\
\nonumber
&+ ik^2M + e^\phi k^2(f\chi - gQ
+ gu) + \Psi e^{2\phi}\lambda f - 2\Psi e^{2\phi}(f^2-g^2) ] \\
\nonumber &+ ik_ak_bk_c(c_3+c_4)[ \Phi - \chi^\prime + i\xi + \Psi
+ iM + e^\phi(f\chi-gQ+gu) ]  \\
\nonumber &+ik_ck_d\epsilon_{abd}(c_4-c_3)[ -ie^\phi g \Phi +
\rho^\prime + e^\phi(g\xi+f\theta) + ie^\phi g \Psi - (f\rho +
gM)e^\phi - ik^2u ] \\
\nonumber &- (c_3+c_4)e^\phi g (\epsilon_{dac}k_bk_d +
\epsilon_{dbc}k_ak_d)[ \Phi - \chi^\prime + i\xi + \Psi + iM +
e^\phi(f\chi - gQ + gu) ] \\
\nonumber &+ \delta_{ac}k_b\{ e^\phi f[ -2(c_3+c_4)i\Psi^\prime ]
+ 2c_3 e^\phi f [ -ie^\phi f \Psi +
 i\sigma
 \\
 \nonumber
 &+ e^\phi(fM-g\rho) ] + 2c_4e^\phi f[ -ie^\phi
f \Phi + M^\prime + e^\phi(f\xi - g\theta)  ]\\
\nonumber &- (c_4-c_3)e^\phi g[ -ie^\phi g \Phi + \rho^\prime +
e^\phi(g\xi + f\theta) + ie^\phi g \Psi - (f\rho+gM)e^\phi - ik^2u
] \} \\
\nonumber &+ \delta_{bc}k_a\{ e^\phi f[ -2(c_3+c_4)i\Psi^\prime ]
+ 2c_4 e^\phi f[ -ie^\phi f \Psi \\
\nonumber
&+
 i\sigma + e^\phi(fM-g\rho) ] + 2c_3e^\phi f[ -ie^\phi
f \Phi + M^\prime \\
\nonumber &+ e^\phi(f\xi - g\theta)  ] + (c_4-c_3)e^\phi g[
-ie^\phi g \Phi + \rho^\prime + e^\phi(g\xi + f\theta) \\
& + ie^\phi g
\Psi - (f\rho+gM)e^\phi - ik^2u ] \} \;.
\end{align*}
Finally we write down the object $S_{ijk}$:

\begin{align*}
S_{a00} = &k_ae^{2\phi}[ -4M(c_3+c_4)(f^2-g^2) + 96c_6\rho
g(\lambda-2f) ] \;,
\\
S_{ab0} = &e^{2\phi}\epsilon_{abc}k_c[ -96c_6g(\lambda-2f)M -
4\rho(c_3+c_4)\lambda f ]\;, 
\\
S_{0ab} = & \delta_{ab}\{ -\frac{2}{3}fe^\phi k^2(c_3+c_4)[-\Phi +
\chi^\prime - i\xi]\\
\nonumber & + 2fe^\phi (c_3+c_4)[ 2e^{2\phi}(f^2-g^2) - 4\Psi
e^{2\phi}(f^2-g^2) - 4e^\phi f\Psi^\prime + 2e^\phi(2f\sigma -
2k^2gu) ]\\
\nonumber & + \frac{10}{3}fe^\phi(c_3+c_4)[ k^2\Psi + ik^2M +
e^\phi k^2(f\chi-gQ+gu) ]\\
\nonumber & + 2(c_3+c_4)e^{2\phi}(f^2-g^2)[
k^2\chi+2\sigma-2e^\phi f\Psi ] - 96c_6ge^{3\phi}(\lambda -
2f)g\Psi \\
\nonumber & + 96c_6ge^\phi[ -2fge^{2\phi} + 4e^{2\phi}fg\Psi +
2e^\phi g\Psi^\prime - 2g\sigma e^\phi + \lambda
ge^{2\phi}(1-\Phi-\Psi) ]\\
\nonumber & + 32c_6ge^\phi k^2[ 2u^\prime + Q^\prime - i\theta -
2i\rho + 2e^\phi(-fQ-gu-g\chi) ] \\
\nonumber & + 48c_6ge^{2\phi}(\lambda - 2f)(Q+u)k^2 \} \\
& + k_ak_b\{
-2fe^\phi(c_3+c_4)[ \Phi - \chi^\prime + i\xi + \Psi + iM +
e^\phi(f\chi - gQ+gu) ]\\
\nonumber & - 2(c_3+c_4)e^{2\phi}(f^2-g^2)\chi +
96c_6ge^{2\phi}(\lambda - 2f)u - 48c_6ge^{2\phi}(\lambda -
2f)(Q+u) \}\\
\nonumber & + \epsilon_{abd}k_d\{ -2fe^\phi(c_4-c_3)[ -ie^\phi
g\Phi +\rho^\prime + e^\phi(g\xi+f\theta) \nonumber \\
& + ie^\phi g\Psi - (f\rho
+gM)e^\phi - ik^2u ]\\
\nonumber
& + 2(c_3+c_4)e^{2\phi}(f^2-g^2)(\rho+\theta) \} \;,
\end{align*}
\begin{align*}
S_{abc} = &\epsilon_{abc}[-96c_6fge^{3\phi}(\lambda - 2f) +
12ge^{3\phi}c_5\lambda f - 12ge^\phi c_5(\Psi^{\prime\prime} +
e^{2\phi}\lambda f(\Psi+\Phi) - \sigma^\prime)\\
\nonumber
& - 96c_6fe^{2\phi}g2\Psi^\prime + \Psi\{
-96c_6f^2e^{3\phi}6g + 96c_6fe^{3\phi}\lambda 2g -
12c_5e^{3\phi}\lambda fg - 4ge^\phi c_5k^2 \}\\
\nonumber
& - \{ ik^2M + e^\phi k^2(f\chi - gQ+gu) \}4ge^\phi c_5
+ 96c_6fe^\phi\{ 2g\sigma e^\phi + \lambda g e^{2\phi}\Phi \}\\
\nonumber
& - 32c_6fe^\phi k^2\{ 2u^\prime + Q^\prime - i\theta -
2i\rho + 2e^\phi(-fQ-fu-g\chi) \}\\
\nonumber
& - 48c_6ge^{2\phi}(\lambda - 2f)(k^2\chi + 2\sigma) +
8ge^\phi k^2c_5\{ -\Phi + \chi^\prime - i\xi \} ]\\
\nonumber
& + [\delta_{ac}k_b - \delta_{bc}k_a][
4fe^\phi(c_3+c_4)i\Psi^\prime - 4fe^\phi c_4\{ -ie^\phi f\Psi +
i\sigma + e^\phi (fM-g\rho) \}\\
\nonumber
& - 4fe^\phi c_3\{ -ie^\phi f\Phi + M^\prime +
e^\phi(f\xi-g\theta) \} \\
& + 2(c_3+c_4)e^{2\phi}\lambda f(\xi+M) -
48c_6ge^{2\phi}(\lambda - 2f)(\rho+\theta)\\
\nonumber
& - 2ge^\phi(c_4-c_3)\{ -ie^\phi g\Phi + \rho^\prime +
e^\phi(g\xi +f\theta) + ie^\phi g\Psi - (f\rho+gM)e^\phi - ik^2u
\} ]\\
\nonumber
& + \epsilon_{abd}k_ck_d[ 48c_6ge^{2\phi}(\lambda -
2f)\chi - 4(c_3+c_4)e^{2\phi}\lambda fu ] \\
\nonumber
&+ [\epsilon_{bfc}k_ak_f - \epsilon_{afc}k_bk_f][
2(c_3+c_4)e^{2\phi}\lambda f (Q+u) \\ \nonumber &+
2ge^\phi(c_3+c_4)\{ \Phi - \chi^\prime + i\xi + \Psi + iM +
e^\phi(f\chi - gQ+ gu) \} ] \;.
\end{align*}

\newpage

\section*{Appendix C}

Here we write the 14 scalar field equations. We use the notations \eqref{all_203}, \eqref{all_204} and \eqref{all_205} and recall the relation \eqref{all_13}.
Equations obtained from \eqref{all_21} are as follows.

$(00)$-component:
\begin{align}
{\cal G}^{(00)} \equiv & -96c_6k^2\Lambda {\mathfrak g}u^\prime - 48c_6k^2\Lambda {\mathfrak g} Q^\prime + (-9\tilde{\beta}{\mathfrak f} - 9\beta \Lambda - 576c_6{\mathfrak g}^2{\mathfrak f}) \Psi^\prime     \nonumber
\\
& + (576c_6{\mathfrak g}^2{\mathfrak f} + 9\tilde{\beta}{\mathfrak f})\sigma + (144c_6{\mathfrak g}^2\Lambda^2 + 9\beta\Lambda^2)\Phi \nonumber \\
& + [k^2(3\tilde{\beta} - 3\beta) - 9\tilde{\beta}({\mathfrak f}^2-{\mathfrak g}^2) + 144c_6{\mathfrak g}^2(\Lambda^2-8{\mathfrak f}^2)]\Psi     \nonumber
\\
& + (-6k^2\tilde{\beta}{\mathfrak g} + 384k^2c_6{\mathfrak f}^2{\mathfrak g})u + \frac{1}{2} (-6k^2\tilde{\beta}{\mathfrak g} + 384k^2c_6{\mathfrak f}^2{\mathfrak g})Q + 3ik^2\tilde{\beta}M    \nonumber
\\
& + (3k^2\tilde{\beta}{\mathfrak f} + 192k^2c_6{\mathfrak g}^2{\mathfrak f})\chi + 192ik^2c_6{\mathfrak f}{\mathfrak g}\rho + 48ik^2c_6\Lambda {\mathfrak g}\theta = 0
\tag{c1}  \label{c1}
\end{align}

$(a0)$-component:
\begin{align}
{\cal G}^{(a0)} \equiv & -i(3\tilde{\beta} + 4\Lambda {\mathfrak f} c_3)M^\prime -i [2\Lambda {\mathfrak g}(c_4-c_3) - 96c_6{\mathfrak g}(\Lambda-2{\mathfrak f})]\rho^\prime + (3\beta - 3\tilde{\beta})\Psi^\prime    \nonumber
\\
& + [-6{\mathfrak g}^2\Lambda c_5 + 4\Lambda c_3({\mathfrak f}^2-{\mathfrak g}^2)]\Psi     -i [6{\mathfrak g}^2\Lambda c_5 - 4\Lambda({\mathfrak f}^2-{\mathfrak g}^2)c_3]M \nonumber
\\
& + [-3\beta\Lambda - 3\tilde{\beta}{\mathfrak f} + 96c_6{\mathfrak g}^2(\Lambda - 2{\mathfrak f}) + 4\Lambda({\mathfrak g}^2-{\mathfrak f}^2)c_3 + 6\Lambda {\mathfrak g}^2c_5]\Phi \nonumber
\\
& - 4\Lambda {\mathfrak f}c_3\sigma -i (8\Lambda {\mathfrak f}{\mathfrak g}c_3 + 6\Lambda {\mathfrak f}{\mathfrak g}c_5)\rho + (6\Lambda {\mathfrak g}k^2c_5 + 4\Lambda {\mathfrak g}k^2c_3)u      \nonumber
 \\
& -i [4\Lambda({\mathfrak f}^2-{\mathfrak g}^2)c_3 - 6{\mathfrak g}^2\Lambda c_5 - 96c_6\Lambda(\Lambda - 2{\mathfrak f}) + 3\tilde{\beta}{\mathfrak f}]\xi    \nonumber
  \\
& -i [-8\Lambda {\mathfrak f}{\mathfrak g}c_3 - 6\Lambda {\mathfrak f}{\mathfrak g}c_5 - 96c_6{\mathfrak f}{\mathfrak g}(\Lambda-2{\mathfrak f}) - 3\tilde{\beta}{\mathfrak g}]\theta=0
\tag{c2}   \label{c2}
\end{align}

$(0a)$-component:
\begin{align}
{\cal G}^{(0a)} \equiv & (3\beta - 3\tilde{\beta})\Psi^\prime- 4ic_3({\mathfrak f}^2-{\mathfrak g}^2)M^\prime - 4i(c_4-c_3){\mathfrak f}{\mathfrak g}\rho^\prime    \nonumber
\\
& + i(12c_3{\mathfrak g}^2{\mathfrak f} + 12{\mathfrak g}^2{\mathfrak f}c_5 - 4c_3{\mathfrak f}^3)\xi -i (-12c_3{\mathfrak g}{\mathfrak f}^2 - 12c_5{\mathfrak g}{\mathfrak f}^2 + 4c_3{\mathfrak g}^3)\theta  \nonumber
\\
& -i [3\tilde{\beta}{\mathfrak f} - 96c_6{\mathfrak g}^2(\Lambda - 2{\mathfrak f}) + 12c_3{\mathfrak g}^2{\mathfrak f} + 12c_5{\mathfrak g}^2{\mathfrak f} - 4c_3{\mathfrak f}^3]M    \nonumber
\\
& -i [-3\tilde{\beta}{\mathfrak g} - 96c_6{\mathfrak g}{\mathfrak f}(\Lambda - 2{\mathfrak f}) + 12c_3{\mathfrak g}{\mathfrak f}^2 + 12c_5{\mathfrak g}{\mathfrak f}^2 - 4c_3{\mathfrak g}^3]\rho    \nonumber
\\
& + [-96c_6k^2{\mathfrak g}(\Lambda - 2{\mathfrak f}) - 4(c_4-c_3){\mathfrak g}{\mathfrak f}k^2]u + [3\tilde{\beta} - 4c_3({\mathfrak f}^2-{\mathfrak g}^2)]\sigma    \nonumber
\\
& + [-3\tilde{\beta}{\mathfrak f} + 96c_6{\mathfrak g}^2(\Lambda - 2{\mathfrak f}) - 12c_3{\mathfrak g}^2{\mathfrak f} - 12c_5{\mathfrak g}^2{\mathfrak f} + 4c_3{\mathfrak f}^3]\Psi   \nonumber
\\
& + [-3\beta\Lambda - 4c_3{\mathfrak f}({\mathfrak f}^2-{\mathfrak g}^2) - 4{\mathfrak g}^2{\mathfrak f}(c_4-c_3)]\Phi = 0
\tag{c3}    \label{c3}
\end{align}

$(ab)$-component, $k_{\ta}k_{b}$:
\begin{align}
{\cal G}^{(k\otimes k)} \equiv & -(48c_6\Lambda {\mathfrak g} - 96c_6{\mathfrak f}{\mathfrak g})Q^\prime - [-3c_5\Lambda {\mathfrak f} - \frac{3}{2}\tilde{\beta} - 3c_5({\mathfrak f}^2-{\mathfrak g}^2)]\chi^\prime  \nonumber
\\
&- (96c_6{\mathfrak f}{\mathfrak g}   -48c_6\Lambda {\mathfrak g})u^\prime - [-\frac{3}{2}\beta + \frac{3}{2}\tilde{\beta} + 3c_5({\mathfrak g}^2-{\mathfrak f}^2-\Lambda {\mathfrak f})]\Psi    \nonumber
\\
 & - [3c_5\Lambda {\mathfrak f} + \frac{3}{2}\tilde{\beta} - \frac{3}{2}\beta + 3c_5({\mathfrak f}^2-{\mathfrak g}^2)]\Phi - (-48ic_6\Lambda {\mathfrak g} + 96ic_6{\mathfrak f}{\mathfrak g})\theta \nonumber \\
 & - (-48ic_6\Lambda {\mathfrak g} + 96ic_6{\mathfrak f}{\mathfrak g})\rho - [3ic_5({\mathfrak f}^2-{\mathfrak g}^2+\Lambda {\mathfrak f}) + \frac{3}{2}i\tilde{\beta}]\xi   \nonumber
 \\
& - (-3c_5{\mathfrak f}^3 - 3c_5\Lambda {\mathfrak f}^2 - 48c_6{\mathfrak g}^2\Lambda + 96c_6{\mathfrak g}^2{\mathfrak f} + 3c_5{\mathfrak g}^2{\mathfrak f} + \frac{3}{2}{\mathfrak f}\tilde{\beta})\chi    \nonumber
\\
& - (3c_5{\mathfrak f}^2{\mathfrak g} + 3c_5\Lambda {\mathfrak f}{\mathfrak g} - 3c_5{\mathfrak g}^3 - 48c_6\Lambda {\mathfrak f}{\mathfrak g} + 96c_6{\mathfrak f}^2{\mathfrak g} - \frac{3}{2}{\mathfrak g}\tilde{\beta})Q   \nonumber
\\
& - [-3ic_5({\mathfrak f}^2-{\mathfrak g}^2+\Lambda {\mathfrak f}) + \frac{3}{2}i\tilde{\beta}]M   \nonumber
\\
& - u(-96c_6{\mathfrak g}{\mathfrak f}^2 - 3c_5{\mathfrak f}^2{\mathfrak g} + 3c_5{\mathfrak g}^3 + 48c_6\Lambda {\mathfrak f}{\mathfrak g} - 3c_5\Lambda {\mathfrak f}{\mathfrak g} + \frac{3}{2}{\mathfrak g}\tilde{\beta})=0
\tag{c4}    \label{c4}
\end{align}

$(ab)$-component, $\delta_{ab}$:
\begin{align}
{\cal G}^{(\delta)} \equiv & (-3\tilde{\beta} +3\beta)\Psi^{\prime\prime} + (-16c_6\Lambda {\mathfrak g}k^2 - 32k^2c_6{\mathfrak f}{\mathfrak g})u^\prime + (-32c_6\Lambda {\mathfrak g}k^2 + 32k^2c_6{\mathfrak f}{\mathfrak g})Q^\prime    \nonumber
 \\
 & + (-192c_6{\mathfrak f}{\mathfrak g}^2 + 6\beta\Lambda + 3{\mathfrak f}\tilde{\beta})\Psi^\prime + (c_5k^2\Lambda {\mathfrak f} + \frac{3}{2}k^2\tilde{\beta} - c_5k^2{\mathfrak g}^2 + c_5k^2{\mathfrak f}^2)\chi^\prime + 3\tilde{\beta}\sigma^\prime    \nonumber
 \\
& + [ -ic_5k^2({\mathfrak f}^2-{\mathfrak g}^2+\Lambda {\mathfrak f}) -\frac{3}{2}ik^2\tilde{\beta}]\xi + (32ik^2c_6{\mathfrak f}{\mathfrak g} + 16ik^2c_6\Lambda {\mathfrak g})\rho + (32ik^2c_6\Lambda {\mathfrak g}     \nonumber
 \\
& + [-c_5k^2({\mathfrak f}^2-{\mathfrak g}^2+\Lambda {\mathfrak f}) + 48c_6\Lambda^2{\mathfrak g}^2 - 3\tilde{\beta}\Lambda {\mathfrak f} - 9\beta\Lambda^2 - \frac{3}{2}k^2\tilde{\beta} + \frac{3}{2}k^2\beta]\Phi    \nonumber
 \\
& + [-3\tilde{\beta}\Lambda {\mathfrak f} + 48c_6\Lambda^2{\mathfrak g}^2 + c_5k^2({\mathfrak f}^2-{\mathfrak g}^2+\Lambda {\mathfrak f}) - 384c_6{\mathfrak f}^2{\mathfrak g}^2 + \frac{3}{2}k^2\beta  \nonumber
\\
&+ 3\tilde{\beta}{\mathfrak f}^2 - \frac{3}{2}k^2\tilde{\beta} - 3\tilde{\beta}{\mathfrak g}^2]\Psi  - 32ik^2c_6 {\mathfrak f}{\mathfrak g})\theta  + (-3\tilde{\beta}{\mathfrak f} + 192c_6{\mathfrak g}^2{\mathfrak f})\sigma     \nonumber
\\
 & + (c_5k^2\Lambda {\mathfrak f}^2 - c_5k^2{\mathfrak g}^2{\mathfrak f} - \frac{3}{2}k^2{\mathfrak f}\tilde{\beta} + c_5k^2{\mathfrak f}^3 +32k^2c_6{\mathfrak f}{\mathfrak g}^2 + 16k^2c_6\Lambda {\mathfrak g}^2)\chi     \nonumber
 \\
& + (-c_5k^2{\mathfrak g}^3 + 160k^2c_6{\mathfrak f}^2{\mathfrak g} + c_5k^2{\mathfrak f}^2{\mathfrak g} + c_5k^2\Lambda {\mathfrak f}{\mathfrak g} - 16k^2c_6\Lambda {\mathfrak g}{\mathfrak f} + \frac{3}{2}k^2{\mathfrak g}\tilde{\beta})u   \nonumber
 \\
& + [32k^2c_6{\mathfrak f}^2{\mathfrak g} + 16k^2c_6\Lambda {\mathfrak g}{\mathfrak f} - c_5k^2\Lambda {\mathfrak f}{\mathfrak g} + c_5k^2{\mathfrak g}^3 + \frac{3}{2}k^2{\mathfrak g}\tilde{\beta} - c_5k^2{\mathfrak f}^2{\mathfrak g}]Q    \nonumber
 \\
& + [ic_5k^2({\mathfrak f}^2-{\mathfrak g}^2+\Lambda {\mathfrak f}) - \frac{3}{2}ik^2\tilde{\beta}]M = 0
\tag{c5}    \label{c5}
\end{align}

$(ab)$-component,  $\epsilon_{abc}k_{c}$:
\begin{align}
{\cal G}^{(\epsilon k)}  \equiv & -i(-3c_5{\mathfrak f}^2 - 3c_5\Lambda {\mathfrak f} - 2c_3\Lambda {\mathfrak f} - 2c_3{\mathfrak g}^2 - \frac{3}{2}\tilde{\beta} + 2c_3{\mathfrak f}^2 + 3c_5{\mathfrak g}^2)\rho^\prime    \nonumber
\\
& -i (4c_3{\mathfrak f}{\mathfrak g} + 12c_5{\mathfrak f}{\mathfrak g} + 96c_6{\mathfrak f}{\mathfrak g} - 48c_6\Lambda {\mathfrak g} - 2c_3\Lambda {\mathfrak g})M^\prime     \nonumber
\\
& + (-12c_5{\mathfrak f}{\mathfrak g} - 6c_5\Lambda {\mathfrak g})\Psi^\prime    \nonumber
\\
& -i (-\frac{3}{2}{\mathfrak g}\tilde{\beta} + 3c_5{\mathfrak g}^3 + 9c_5{\mathfrak f}^2{\mathfrak g} - 2c_3{\mathfrak g}^3 + 6c_3{\mathfrak f}^2{\mathfrak g} \nonumber
\\
&- 3c_5\Lambda {\mathfrak f}{\mathfrak g} - 48c_6\Lambda {\mathfrak f}{\mathfrak g} + 96c_6{\mathfrak g}{\mathfrak f}^2 - 4c_3\Lambda {\mathfrak f}{\mathfrak g})\xi     \nonumber
\\
& + (\frac{3}{2}k^2\tilde{\beta} + 2k^2c_3\Lambda {\mathfrak f} + 3k^2c_5{\mathfrak g}^2 + 2k^2c_3{\mathfrak g}^2 - 3k^2c_5{\mathfrak f}^2 - 3k^2c_5\Lambda {\mathfrak f} - 2k^2c_3{\mathfrak f}^2)u    \nonumber
\\
 & + (-96c_6{\mathfrak f}{\mathfrak g} + 2c_3\Lambda {\mathfrak g} + 48c_6\Lambda {\mathfrak g} + 6c_5\Lambda {\mathfrak g} - 4c_3{\mathfrak f}{\mathfrak g})\sigma    \nonumber
 \\
& -i (-3c_5{\mathfrak f}^3 + 48c_6{\mathfrak g}^2\Lambda + 2c_3{\mathfrak f}^3 - 6c_3{\mathfrak g}^2{\mathfrak f} - 3c_5\Lambda {\mathfrak f}^2 - 9c_5{\mathfrak g}^2{\mathfrak f} - 2c_3\Lambda {\mathfrak f}^2  \nonumber \\
& - \frac{3}{2}{\mathfrak f}\tilde{\beta} + 2c_3{\mathfrak g}^2\Lambda - 96c_6{\mathfrak f}{\mathfrak g}^2)\theta    \nonumber
\\
& + (-2c_3{\mathfrak f}^2{\mathfrak g} + 3c_5{\mathfrak f}^2{\mathfrak g} - 96c_6{\mathfrak g}{\mathfrak f}^2 + 3c_5{\mathfrak f}{\mathfrak g}\Lambda + \frac{3}{2}{\mathfrak g}\beta + 2c_3\Lambda {\mathfrak f}{\mathfrak g} - 4c_3{\mathfrak f}^2{\mathfrak g}    \nonumber
\\
& + 48c_6\Lambda {\mathfrak f}{\mathfrak g} + 2c_3\Lambda {\mathfrak f}{\mathfrak g} - 12c_5{\mathfrak f}^2{\mathfrak g} - 3c_5{\mathfrak g}^3 + 2c_3{\mathfrak g}^3)\Phi    \nonumber
\\
& + (-3c_5\Lambda {\mathfrak f}{\mathfrak g} + 3c_5{\mathfrak f}^2{\mathfrak g} - 3c_5{\mathfrak g}^3 - 2c_3{\mathfrak g}^3 - \frac{3}{2}{\mathfrak g}\tilde{\beta} \nonumber \\
& + 96c_6{\mathfrak g}{\mathfrak f}^2 + 6c_3{\mathfrak f}^2{\mathfrak g} - 4c_3\Lambda {\mathfrak f}{\mathfrak g} - 48c_6\Lambda {\mathfrak f}{\mathfrak g})\Psi    \nonumber
\\
 & -i (-6c_3{\mathfrak f}^2{\mathfrak g} + 48c_6\Lambda {\mathfrak f}{\mathfrak g} + \frac{3}{2}{\mathfrak g}\tilde{\beta} - 3c_5{\mathfrak f}^2{\mathfrak g} + 3c_5\Lambda {\mathfrak f}{\mathfrak g} \nonumber \\
 & - 96c_6{\mathfrak g}{\mathfrak f}^2 + 4c_3\Lambda {\mathfrak f}{\mathfrak g} + 3c_5{\mathfrak g}^3 + 2c_3{\mathfrak g}^3)M   \nonumber
 \\
 & -i (-6c_5{\mathfrak g}^2\Lambda + 3c_5{\mathfrak g}^2{\mathfrak f} + 2c_3\Lambda {\mathfrak f}^2 - 3c_5\Lambda {\mathfrak f}^2 - 48c_6{\mathfrak g}^2\Lambda + 96c_6{\mathfrak f}{\mathfrak g}^2 \nonumber
 \\
 & + 6c_3{\mathfrak g}^2{\mathfrak f} - 2c_3{\mathfrak g}^2\Lambda + \frac{3}{2}\tilde{\beta}{\mathfrak f} - 3c_5{\mathfrak f}^3 - 2c_3{\mathfrak f}^3)\rho = 0
\tag{c6}     \label{c6}
\end{align}

Equations obtained from \eqref{all_22} are as follows.

$(0a0)$-component:
\begin{align}
{\cal T}^{(0a0)} \equiv & -2c_5k^2\chi^\prime + i( 4c_4{\mathfrak g} + 6c_5{\mathfrak g} )\rho^\prime + 4c_3i{\mathfrak f}M^\prime + 2c_5k^2{\mathfrak g}Q \nonumber
\\
& + i(6c_5{\mathfrak g}^2 + 2k^2c_5 + 4c_3{\mathfrak f}^2 + 4c_4{\mathfrak g}^2)\xi + ( 16c_5k^2{\mathfrak g} + 4{\mathfrak g}c_4k^2 )u \nonumber
\\
& + 4c_3{\mathfrak f}\sigma + i( 8c_4{\mathfrak g}{\mathfrak f} + 18c_5{\mathfrak g}{\mathfrak f} )\theta -i ( 96\Lambda c_6{\mathfrak g} - 192c_6{\mathfrak f}{\mathfrak g} + 8c_4{\mathfrak g}{\mathfrak f} + 6c_5{\mathfrak g}{\mathfrak f} )\rho \nonumber
\\
& + ( 4c_3{\mathfrak f}^2 + 4c_4{\mathfrak g}^2 + 2k^2c_5 + 6c_5{\mathfrak g}^2 )\Phi - 2ik^2c_5{\mathfrak f}\chi  \nonumber
\\
& + ( -2c_5k^2 - 4c_4{\mathfrak g}^2 - 4c_3{\mathfrak f}^2 - 18c_5{\mathfrak g}^2 )\Psi  \nonumber
\\
& -i ( 4c_3{\mathfrak f}^2 + 18c_5{\mathfrak g}^2 - 3\tilde{\beta} + 4c_4{\mathfrak g}^2 + 2c_5k^2 )M = 0
\tag{c7}     \label{c7}
\end{align}

$(ab0)$-component:
\begin{align}
{\cal T}^{(ab0)} \equiv & -32c_6k^2u^\prime - 16c_6k^2Q^\prime - ( 96c_6{\mathfrak g} + 12c_5{\mathfrak g} )\Psi^\prime -i ( 6c_5{\mathfrak f} + 4c_4{\mathfrak f} )\rho^\prime  \nonumber
 \\
& + 4ic_3{\mathfrak g}M^\prime + 32c_6k^2{\mathfrak f}Q +i ( 96\Lambda c_6{\mathfrak g} + 6c_5{\mathfrak g}{\mathfrak f} - 192c_6{\mathfrak f}{\mathfrak g} + 8c_4{\mathfrak g}{\mathfrak f} )M   \nonumber
\\
& + ( 48c_6\Lambda {\mathfrak g} + 8c_4{\mathfrak g}{\mathfrak f} - 192c_6{\mathfrak f}{\mathfrak g} + 6c_5{\mathfrak g}{\mathfrak f} )\Psi + 32c_6k^2{\mathfrak g}\chi    \nonumber
\\
& -i ( 8c_4{\mathfrak g}{\mathfrak f} + 18c_5{\mathfrak g}{\mathfrak f} )\xi - ( -64c_6k^2{\mathfrak f} + 4{\mathfrak f}k^2c_4 + 6{\mathfrak f}k^2c_5 )u  \nonumber
\\
& + ( 96c_6{\mathfrak g} - 4c_4{\mathfrak g} )\sigma +i ( 16c_6k^2 - 4c_4{\mathfrak f}^2 - 4c_3{\mathfrak g}^2 - 6c_5{\mathfrak f}^2 )\theta  \nonumber
\\
& +i ( 32c_6k^2 + 4c_4{\mathfrak f}^2 + 3\tilde{\beta} + 18c_5{\mathfrak f}^2 + 4c_3{\mathfrak g}^2 )\rho \nonumber \\
& - ( 8c_4{\mathfrak g}{\mathfrak f} + 18c_5{\mathfrak g}{\mathfrak f} - 48c_6\Lambda {\mathfrak g} )\Phi = 0
\tag{c8}     \label{c8}
\end{align}

$(0ab)$-component, $k_{a}k_{b}$:
\begin{align}
{\cal T}^{(k\otimes k)} \equiv & 3c_5\chi^{\prime\prime} + 3c_5{\mathfrak g}Q^\prime - 3c_5\Phi^\prime - 3c_5{\mathfrak g}u^\prime - (2ic_4 +3ic_5)M^\prime - 3ic_5\xi^\prime + 3c_5\Psi^\prime   \nonumber
\\
& - ( -3c_5\Lambda {\mathfrak f} + 3c_5{\mathfrak f}^2 + \frac{3}{2}\tilde{\beta} )\chi - (3ic_5{\mathfrak f} + 2ic_4{\mathfrak f})\xi   \nonumber
\\
 & - (3c_5{\mathfrak f} + 2c_4{\mathfrak f})\Phi - ( 2c_4 + 6c_5 )\sigma + 2ic_3{\mathfrak g}\rho  \nonumber
\\
& - ( 3c_5\Lambda {\mathfrak g} - 96c_6{\mathfrak f} {\mathfrak g} + 3c_5{\mathfrak f} {\mathfrak g} + 48c_6\Lambda {\mathfrak g} )u - ( -3c_5{\mathfrak f} - 2c_4{\mathfrak f} )\Psi + 2ic_4{\mathfrak g}\theta  \nonumber
\\
 & - ( -2ic_4{\mathfrak f} - 3ic_5{\mathfrak f} )M - ( -48\Lambda c_6 {\mathfrak g} + 96c_6{\mathfrak f}{\mathfrak g} - 3c_5\Lambda {\mathfrak g} - 3c_5{\mathfrak f}{\mathfrak g} )Q = 0
\tag{c9}     \label{c9}
\end{align}

$(0ab)$-component, $\delta_{ab}$:
\begin{align}
{\cal T}^{(\delta)} \equiv & c_5k^2\chi^{\prime\prime} + ( 5k^2c_5{\mathfrak g} + 32k^2c_6{\mathfrak g} )Q^\prime - c_5k^2\Phi^\prime + ( c_5k^2 + 192c_6{\mathfrak g}^2 )\Psi^\prime   \nonumber
 \\
& -ik^2c_5\xi^\prime + (-5ik^2c_5 - 2ik^2c_4)M^\prime + (64k^2c_6{\mathfrak g} + 7k^2c_5{\mathfrak g})u^\prime \nonumber
\\
& + ( 5ik^2c_5{\mathfrak f} + 2ik^2c_4{\mathfrak f} )M   +( -160k^2c_6{\mathfrak g}{\mathfrak f} + 5k^2c_5\Lambda {\mathfrak g} + 48k^2c_6\Lambda {\mathfrak g} + k^2c_5{\mathfrak f}{\mathfrak g} )Q   \nonumber
\\
& + ( -32ik^2c_6{\mathfrak g} + 2ik^2c_4{\mathfrak g} )\theta +( 48k^2c_6\Lambda {\mathfrak g} - k^2c_5{\mathfrak f}{\mathfrak g} - 224k^2c_6{\mathfrak f}{\mathfrak g} + 7k^2c_5\Lambda {\mathfrak g} )u  \nonumber
\\
& + (-5ik^2c_5{\mathfrak f} - 2ik^2c_4{\mathfrak f})\xi +( 2k^2c_3 - 3\tilde{\beta} - 192c_6{\mathfrak g}^2 )\sigma \nonumber \\
&  + ( 2k^2c_4{\mathfrak f} + 5k^2c_5{\mathfrak f} + 576c_6{\mathfrak g}^2{\mathfrak f} + 3{\mathfrak f}\tilde{\beta} - 24c_5\Lambda {\mathfrak g}^2 - 192c_6\Lambda {\mathfrak g}^2 )\Psi   \nonumber
\\
& +( - 96c_6\Lambda {\mathfrak g}^2 - 5k^2c_5{\mathfrak f} - 2k^2c_4{\mathfrak f} - 12c_5\Lambda {\mathfrak f}^2 + 3\Lambda \tilde{\beta} )\Phi \nonumber \\
&  + ( -64ik^2c_6{\mathfrak g} + 2ik^2c_3{\mathfrak g} )\rho  \nonumber
\\
&  + ( -k^2c_5{\mathfrak f}^2 - \frac{3}{2}k^2\tilde{\beta} + k^2c_5\Lambda {\mathfrak f} - 64k^2c_6{\mathfrak g}^2 )\chi = 0
\tag{c10}     \label{c10}
\end{align}

$(0ab)$-component, $\epsilon_{abc}k_{c}$:
\begin{align}
{\cal T}^{(\epsilon k)} \equiv & i(2c_4+3c_5)\rho^{\prime\prime} - 3ic_5{\mathfrak g}M^\prime - 16k^2c_6Q^\prime +i ( 2c_4{\mathfrak f} + 3c_5{\mathfrak f} )\theta^\prime  \nonumber
\\
& + ( 3k^2c_5 + 2k^2c_4 - 32k^2c_6 )u^\prime + ( 3c_5{\mathfrak g} + 2c_4{\mathfrak g} )\Phi^\prime   \nonumber
\\
& +i ( 2c_4{\mathfrak g}+3c_5{\mathfrak g} )\xi^\prime + ( -2c_4{\mathfrak g} - 9c_5{\mathfrak g} - 96c_6{\mathfrak g} )\Psi^\prime   \nonumber \\
& -i ( -3c_5{\mathfrak g}{\mathfrak f} - 48c_6\Lambda {\mathfrak g} - 4c_4{\mathfrak g}{\mathfrak f} + 96c_6{\mathfrak g}{\mathfrak f} - 2c_4\Lambda {\mathfrak g} - 3c_5\Lambda {\mathfrak g} )\xi    \nonumber
\\
 & + ( 2{\mathfrak f}k^2c_4 + 64k^2c_6{\mathfrak f} + 3k^2{\mathfrak f}c_5 )u + ( 96c_6{\mathfrak g} + 6c_5{\mathfrak g} + 2c_4{\mathfrak g} )\sigma  \nonumber
 \\
& -i ( -9c_5{\mathfrak f}^2 - 2c_4\Lambda {\mathfrak f} + 6c_5{\mathfrak g}^2 - 2c_4{\mathfrak f}^2 - 16c_6k^2 - \frac{3}{2}\tilde{\beta} - 3c_5\Lambda {\mathfrak f} + 2c_4{\mathfrak g}^2 )\theta   \nonumber
\\
 & -i ( -2c_4{\mathfrak g}^2 - 32c_6k^2 + 9c_5\Lambda {\mathfrak f} - 6c_5{\mathfrak g}^2 - \frac{3}{2}\tilde{\beta} + 2c_4\Lambda {\mathfrak f} + 2c_4{\mathfrak f}^2 + 3c_5{\mathfrak f}^2 )\rho  \nonumber
\\
& + ( 3c_5\Lambda {\mathfrak g} + 2c_4{\mathfrak f}{\mathfrak g} + 2c_4\Lambda {\mathfrak g} + 2c_4{\mathfrak g}{\mathfrak f} + 48c_6\Lambda {\mathfrak g} + 3c_5{\mathfrak f}{\mathfrak g} )\Phi   \nonumber
\\
& + ( -4c_4{\mathfrak g}{\mathfrak f} - 2c_4\Lambda {\mathfrak g} - 192c_6{\mathfrak f}{\mathfrak g} + 48c_6\Lambda {\mathfrak g} - 9c_5{\mathfrak f}{\mathfrak g} - 3c_5\Lambda {\mathfrak g} )\Psi   \nonumber
\\
& + 32c_6k^2{\mathfrak g}\chi -i ( 4c_4{\mathfrak g}{\mathfrak f} + 2c_4\Lambda {\mathfrak g} + 3c_5\Lambda {\mathfrak g} + 9c_5{\mathfrak g}{\mathfrak f} + 96c_6{\mathfrak f}{\mathfrak g} - 48c_6\Lambda {\mathfrak g} )M  \nonumber
\\
& + 32k^2c_6{\mathfrak f}Q = 0
\tag{c11}     \label{c11}
\end{align}

$(abc)$-component, $\epsilon_{abd}k_{c}k_{d}$:
\begin{align}
{\cal T}^{(\epsilon k \otimes k)} \equiv & -(3ic_5+2ic_4)\rho^\prime + 3c_5{\mathfrak g}\chi^\prime + 2ic_3{\mathfrak g}\xi + 2c_4{\mathfrak g}\Psi + 2ic_4{\mathfrak g}M \nonumber \\
& - (3c_5{\mathfrak g}^2 - \frac{3}{2}\tilde{\beta} + 3k^2c_5 + 2k^2c_4 + 6c_5\Lambda {\mathfrak f})u  \nonumber
\\
 & - (2ic_4{\mathfrak f}+3ic_5{\mathfrak f})\theta + (2ic_4{\mathfrak f} + 3ic_5{\mathfrak f})\rho + 2c_3{\mathfrak g}\Phi   \nonumber
 \\
 & - (48c_6\Lambda {\mathfrak g} - 96c_6{\mathfrak f}{\mathfrak g} + 3c_5{\mathfrak g}{\mathfrak f})\chi  - (\frac{3}{2}\tilde{\beta} - 6c_5\Lambda {\mathfrak f} - 3c_5{\mathfrak g}^2)Q = 0
\tag{c12}     \label{c12}
\end{align}

$(abc)$-component, $\delta_{ac}k_{b} - \delta_{ab}k_{c}$:
\begin{align}
{\cal T}^{(\delta\otimes k)} \equiv & 2ic_3M^{\prime\prime} - 2c_3{\mathfrak f}\Psi^\prime + 2c_3{\mathfrak f}\Phi^\prime + 3ic_5{\mathfrak g}\rho^\prime - 2ic_3{\mathfrak g}\theta^\prime + 2ic_3{\mathfrak f}\xi^\prime + 2c_3\sigma^\prime - k^2c_5\chi^\prime   \nonumber
\\
& -i (-2c_3{\mathfrak f}^2 + \frac{3}{2}\tilde{\beta} - k^2c_5 - 2c_3\Lambda {\mathfrak f} - 2c_4{\mathfrak g}^2 - 9c_5{\mathfrak g}^2 )\xi + (2{\mathfrak g}c_4k^2 + 2{\mathfrak g}c_5k^2)u   \nonumber
 \\
& + 2c_3{\mathfrak f}\sigma -i ( 2c_3\Lambda {\mathfrak g} - 3c_5{\mathfrak f}{\mathfrak g} + 96c_6 {\mathfrak f}{\mathfrak g} - 48c_6\Lambda {\mathfrak g} - 4c_4{\mathfrak g}{\mathfrak f} )\theta   \nonumber
 \\
 & -i (-48c_6\Lambda {\mathfrak g} + 4c_4{\mathfrak g}{\mathfrak f} + 9c_5 {\mathfrak g}{\mathfrak f} + 2c_4\Lambda {\mathfrak g} + 96c_6 {\mathfrak f}{\mathfrak g})\rho \nonumber \\
 & + (2c_3\Lambda {\mathfrak f} + 2c_3{\mathfrak f}^2 - 2c_3{\mathfrak g}^2 + 3c_5{\mathfrak g}^2 + c_5k^2)\Phi  \nonumber
  \\
& + ( -2c_3\Lambda {\mathfrak f} - 2c_4{\mathfrak g}^2 - 2c_3{\mathfrak f}^2 - 3c_5{\mathfrak g}^2 - k^2c_5 )\Psi -  c_5k^2{\mathfrak f}\chi  \nonumber
\\
& -i ( k^2c_5 + \frac{3}{2}\tilde{\beta} + 2c_4{\mathfrak g}^2 + 2c_3{\mathfrak f}^2 + 3c_5{\mathfrak g}^2 + 2c_3\Lambda {\mathfrak f} )M + c_5k^2{\mathfrak g}Q = 0
\tag{c13}     \label{c13}
\end{align}

$(abc)$-component, $\epsilon_{abc}$:
\begin{align}
{\cal T}^{(\epsilon)} \equiv & -32c_6k^2u^{\prime\prime} - 16c_6k^2Q^{\prime\prime} + (-96c_6{\mathfrak g} - 12c_5{\mathfrak g})\Psi^{\prime\prime} + (5k^2c_5{\mathfrak g} + 32k^2c_6{\mathfrak g})\chi^\prime   \nonumber
 \\
& + ( 32ic_6k^2 - 2ik^2c_4 - 3ik^2c_5 )\rho^\prime + ( -48c_6\Lambda {\mathfrak g} - 384c_6{\mathfrak f}{\mathfrak g} )\Psi^\prime + 48c_6\Lambda {\mathfrak g}\Phi^\prime   \nonumber
 \\
 & + 16ik^2c_6\theta^\prime + (12c_5{\mathfrak g}+96c_6{\mathfrak g})\sigma^\prime   + (-8ik^2c_5{\mathfrak g} - 2ik^2c_4{\mathfrak g})\xi    \nonumber
 \\
& + ( 32ik^2c_6{\mathfrak f} - 3ik^2c_5{\mathfrak f} - 2ik^2c_4{\mathfrak f} )\theta + ( 160k^2c_6{\mathfrak g}{\mathfrak f} - k^2c_5{\mathfrak f}{\mathfrak g} - 16k^2\Lambda {\mathfrak g} c_6 )\chi \nonumber
\\
& + ( -3k^4c_5 - 2k^4c_4 + 64k^2c_6\Lambda {\mathfrak f} - \frac{3}{2}k^2\tilde{\beta} - k^2c_5{\mathfrak g}^2 + 6k^2c_5\Lambda {\mathfrak f} + 128k^2c_6{\mathfrak f}^2 )u  \nonumber
 \\
& + (-2k^2c_4{\mathfrak g} - 3k^2c_5{\mathfrak g} + 96c_6\Lambda^2 {\mathfrak g} - 12c_5\Lambda {\mathfrak f} {\mathfrak g} - 5k^2c_5{\mathfrak g} + 96c_6\Lambda {\mathfrak f} {\mathfrak g})\Phi  \nonumber
\\
& + ( 32k^2c_6\Lambda {\mathfrak f} + 64k^2c_6{\mathfrak f}^2 + k^2c_5{\mathfrak g}^2 +6k^2c_5\Lambda {\mathfrak f} - \frac{3}{2}k^2\tilde{\beta} )Q + 384c_6{\mathfrak f}{\mathfrak g}\sigma \nonumber
\\
& + ( 2k^2c_5{\mathfrak g} - 192c_6\Lambda {\mathfrak f} {\mathfrak g} + 3{\mathfrak g}\tilde{\beta} - 576c_6{\mathfrak f}^2{\mathfrak g} + 96c_6\Lambda^2{\mathfrak g} + 2k^2c_4{\mathfrak g} - 24c_5\Lambda {\mathfrak f} {\mathfrak g} )\Psi  \nonumber
 \\
& + ( 2ik^2c_4{\mathfrak g} + 2ik^2c_5{\mathfrak g} )M   + ( 3ik^2c_5{\mathfrak f} + 64ik^2c_6{\mathfrak f} + 2ik^2c_4{\mathfrak f} )\rho = 0
\tag{c14}      \label{c14}
\end{align}

Note that eqs.~\eqref{c5}, \eqref{c9}, \eqref{c10}, \eqref{c11}, \eqref{c13} and \eqref{c14} are second order, while eqs.~\eqref{c1}-\eqref{c4}, \eqref{c6}-\eqref{c8} and \eqref{c12} are first order.

\section*{Appendix D} 
Here we give explicit forms of eqs.~\eqref{all_209} and \eqref{all_210}:
\begin{align}
{\cal D}^{(1)} & \equiv   4ic_3k^2M^\prime      + ( 12c_5{\mathfrak g}k^2 + 96c_6{\mathfrak g}k^2)Q^\prime + 576c_6{\mathfrak g}^2\Psi^\prime + 2( 12c_5{\mathfrak g}k^2 + 96c_6{\mathfrak g}k^2)u^\prime   \nonumber
\\
& + (  4c_3k^2 - 9\tilde{\beta}- 576c_6{\mathfrak g}^2)\sigma  + 4ik^2c_3{\mathfrak f}\xi + ( -192ik^2c_6{\mathfrak g} + 4ik^2c_3{\mathfrak g} )\rho \nonumber \\
&+ ( 12k^2c_5\Lambda {\mathfrak g} + 96c_6k^2\Lambda {\mathfrak g} - 384c_6k^2{\mathfrak g}{\mathfrak f} )Q + 2( 12k^2c_5\Lambda {\mathfrak g} \nonumber \\
& -  4ik^2c_3{\mathfrak f}M + 96c_6k^2\Lambda {\mathfrak g} - 384c_6k^2{\mathfrak g}{\mathfrak f} )u \nonumber \\
&+ ( -96c_6ik^2{\mathfrak g} + 4ik^2c_4{\mathfrak g} )\theta    + (-3k^2\tilde{\beta}  - 192k^2c_6{\mathfrak g}^2)\chi \nonumber
\\
&+ (  4k^2c_3{\mathfrak f} - 36c_5\Lambda {\mathfrak f}^2 - 288c_6\Lambda {\mathfrak g}^2 +9\tbeta \Lambda )\Phi \nonumber \\
& + [ -72c_5\Lambda {\mathfrak g}^2 -  4k^2c_3{\mathfrak f} + 9\tilde{\beta}({\mathfrak f}+\Lambda) - 576c_6{\mathfrak g}^2(\Lambda - 3{\mathfrak f}) ]\Psi  = 0  \;, \tag{d1} \label{d1}
\end{align}

\begin{align}
{\cal D}^{(2)} & \equiv [ 3\tilde{\beta}-12c_5\Lambda {\mathfrak f} ]u^\prime + [ -3\tilde{\beta} + 12c_5\Lambda {\mathfrak f} ]Q^\prime + [ 6c_5\Lambda {\mathfrak g} - 96c_6{\mathfrak g}(\Lambda-2{\mathfrak f}) ]\chi^\prime  \nonumber
\\
& + 8(c_4-c_3)i{\mathfrak g}M^\prime   + 4(c_4-c_3)i{\mathfrak f}\rho^\prime + [ 3i\tilde{\beta} - 8ic_3({\mathfrak f}^2-{\mathfrak g}^2) - 4(c_4-c_3)i{\mathfrak g}^2 ]\theta \nonumber
\\
&  + 8(c_4-c_3){\mathfrak g}\sigma   + [ 96ic_6{\mathfrak g}(\Lambda-2{\mathfrak f}) - 6ic_5\Lambda {\mathfrak g} + 12(c_4-c_3)i{\mathfrak f}{\mathfrak g} ]\xi \nonumber
\\
& + [ -96c_6{\mathfrak g}i(\Lambda-2{\mathfrak f}) - 6ic_5\Lambda {\mathfrak g} - 12i(c_4-c_3){\mathfrak f}{\mathfrak g} ]M  \nonumber
\\
& + [ -3i\tilde{\beta} - 12ic_5\Lambda {\mathfrak f} + 16ic_3({\mathfrak f}^2-{\mathfrak g}^2) + 4(c_4-c_3)i{\mathfrak f}^2 ]\rho \nonumber \\
& + [ -12(c_4-c_3){\mathfrak f}{\mathfrak g} - 6c_5\Lambda {\mathfrak g} ]\Psi + [ 12(c_4-c_3){\mathfrak f}{\mathfrak g} - 6c_5\Lambda {\mathfrak g} ]\Phi \nonumber
\\
& + [ 3\tilde{\beta}{\mathfrak g} - 18c_5\Lambda {\mathfrak f}{\mathfrak g} + 192c_6{\mathfrak f}^2{\mathfrak g} - 192c_6\Lambda^2{\mathfrak g} + 288c_6\Lambda {\mathfrak f}{\mathfrak g} ]\chi \nonumber
\\
& + [ 96c_6{\mathfrak g}^2(\Lambda-2{\mathfrak f}) + 6\Lambda \tilde{\beta} + 3{\mathfrak f}\tilde{\beta} +  4(c_4-c_3){\mathfrak f}k^2 - 6c_5\Lambda {\mathfrak g}^2 - 12c_5\Lambda {\mathfrak f}^2 - 24c_5\Lambda^2{\mathfrak f} ]u \nonumber
\\
& + [ -3{\mathfrak f}\tilde{\beta} - 6\Lambda \tilde{\beta} - 96c_6{\mathfrak g}^2(\Lambda-2{\mathfrak f}) + 6c_5\Lambda {\mathfrak g}^2 + 12c_5\Lambda {\mathfrak f}^2 + 24c_5\Lambda^2 {\mathfrak f} ]Q = 0 \;.  \tag{d2} \label{d2}
\end{align}
These equations are obviously first order.


\vspace*{1cm}

\begin{thebibliography}{20}



\bibitem{1}
G.~Gabadadze,
  ``ICTP lectures on large extra dimensions,''
  hep-ph/0308112.


\bibitem{2}
Y. Fujii, K. Maeda, 
``The Scalar-Tensor Theory of Gravitation'' (Cambridge University Press, 2003).


  \bibitem{3}
V.~A.~Rubakov and P.~G.~Tinyakov,
  ``Infrared-modified gravities and massive gravitons,''
  Phys.\ Usp.\  {\bf 51}, 759 (2008)
  doi:10.1070/PU2008v051n08ABEH006600
  [arXiv:0802.4379 [hep-th]].
  
  
  \bibitem{4}
A.~De Felice and S.~Tsujikawa,
  ``f(R) theories,''
  Living Rev.\ Rel.\  {\bf 13}, 3 (2010)
  doi:10.12942/lrr-2010-3
  [arXiv:1002.4928 [gr-qc]].


\bibitem{5}
K.~Hinterbichler,
  ``Theoretical Aspects of Massive Gravity,''
  Rev.\ Mod.\ Phys.\  {\bf 84}, 671 (2012)
  doi:10.1103/RevModPhys.84.671
  [arXiv:1105.3735 [hep-th]].



\bibitem{6}
T.~Clifton, P.~G.~Ferreira, A.~Padilla and C.~Skordis,
  ``Modified Gravity and Cosmology,''
  Phys.\ Rept.\  {\bf 513}, 1 (2012)
  doi:10.1016/j.physrep.2012.01.001
  [arXiv:1106.2476 [astro-ph.CO]].




\bibitem{HLecture}
F.~W.~Hehl,
\textit{Cosmology and Gravitation,}
``Four Lectures On Poincare Gauge Field Theory,''
  NATO Advanced Study Institutes Series, Vol. 58 (Springer, 1980).


\bibitem{book1}
  M.~Blagojevi\'c,
  ``Gravitation and gauge symmetries,''
  Bristol, UK: IOP (2002).


\bibitem{book3}
M.~Blagojevi\'c and F.~W.~Hehl,
``Gauge Theories of Gravitation : A Reader with Commentaries''  (Imperial College Press, 2013).


\bibitem{book2}
T.~Ortin,
``Gravity and strings'' (Cambridge University Press, 2015).







\bibitem{HS-1}
K.~Hayashi and T.~Shirafuji,
  ``Gravity from Poincar\'e Gauge Theory of the Fundamental Particles. 1. Linear and Quadratic Lagrangians,''
  Prog.\ Theor.\ Phys.\  {\bf 64}, 866 (1980)
  Erratum: [Prog.\ Theor.\ Phys.\  {\bf 65}, 2079 (1981)].
  doi:10.1143/PTP.64.866

\bibitem{HS-2}
K.~Hayashi and T.~Shirafuji,
  ``Gravity From Poincar\'e Gauge Theory of the Fundamental Particles. 3. Weak Field Approximation,''
  Prog.\ Theor.\ Phys.\  {\bf 64}, 1435 (1980)
  Erratum: [Prog.\ Theor.\ Phys.\  {\bf 66}, 741 (1981)].
  doi:10.1143/PTP.64.1435

\bibitem{HS-3}
K.~Hayashi and T.~Shirafuji,
  ``Gravity From Poincar\'e Gauge Theory of the Fundamental Particles. 4. Mass and Energy of Particle Spectrum,''
  Prog.\ Theor.\ Phys.\  {\bf 64}, 2222 (1980).
  doi:10.1143/PTP.64.2222

\bibitem{Sezgin}
E.~Sezgin and P.~van Nieuwenhuizen,
  ``New Ghost Free Gravity Lagrangians with Propagating Torsion,''
  Phys.\ Rev.\ D {\bf 21}, 3269 (1980).
  doi:10.1103/PhysRevD.21.3269





\bibitem{H2}
P.~Baekler, F.~W.~Hehl and J.~M.~Nester,
``Poincare gauge theory of gravity: Friedman cosmology with even and odd parity modes. Analytic part,''
  Phys.\ Rev.\ D {\bf 83}, 024001 (2011)
  doi:10.1103/PhysRevD.83.024001
  [arXiv:1009.5112 [gr-qc]].



\bibitem{42}
A.~V.~Minkevich,
  ``Generalized Cosmological Friedmann Equations And The De Sitter Solution,''
  Phys.\ Lett.\ A {\bf 95}, 422 (1983).
  doi:10.1016/0375-9601(83)90309-2
  

\bibitem{49} 
  K.~F.~Shie, J.~M.~Nester and H.~J.~Yo,
  ``Torsion Cosmology and the Accelerating Universe,''
  Phys.\ Rev.\ D {\bf 78}, 023522 (2008)
  doi:10.1103/PhysRevD.78.023522
  [arXiv:0805.3834 [gr-qc]].
  

\bibitem{44}
A.~V.~Minkevich,
  ``De Sitter spacetime with torsion as physical spacetime in the vacuum,''
  Mod.\ Phys.\ Lett.\ A {\bf 26}, 259 (2011)
  doi:10.1142/S0217732311034797
  [arXiv:1002.0538 [gr-qc]].

  
\bibitem{50}
  X.~C.~Ao and X.~Z.~Li,
  ``Torsion Cosmology of Poincar\'e gauge theory and the constraints of its parameters via SNeIa data,''
  JCAP {\bf 1202} (2012) 003
  doi:10.1088/1475-7516/2012/02/003
  [arXiv:1111.2385 [gr-qc]].
    


\bibitem{47}
  G.~Chee and Y.~Guo,
  ``Exact de Sitter solutions in quadratic gravitation with torsion,''
  Class.\ Quant.\ Grav.\  {\bf 29} (2012) 235022
  doi:10.1088/0264-9381/29/23/235022
  [arXiv:1205.5419 [gr-qc]].
    

\bibitem{51}
 C.~Q.~Geng, C.~C.~Lee and H.~H.~Tseng,
  ``Scalar-Torsion Cosmology in the Poincar\'e Gauge Theory of Gravity,''
  JCAP {\bf 1211} (2012) 013
  doi:10.1088/1475-7516/2012/11/013
  [arXiv:1207.0579 [gr-qc]].
 


\bibitem{46}
A.~S.~Garkun, V.~I.~Kudin and A.~V.~Minkevich,
  ``To theory of asymptotically stable accelerating Universe in Riemann-Cartan spacetime,''
  JCAP {\bf 1412}, no. 12, 027 (2014)
  doi:10.1088/1475-7516/2014/12/027
  [arXiv:1410.0460 [gr-qc]].


\bibitem{48}
  J.~Lu and G.~Chee,
  ``Cosmology in Poincar\'e gauge gravity with a pseudoscalar torsion,''
  JHEP {\bf 1605} (2016) 024
  doi:10.1007/JHEP05(2016)024
  [arXiv:1601.03943 [gr-qc]].



\bibitem{last}
V.~Nikiforova, S.~Randjbar-Daemi and V.~Rubakov,
  ``Self-accelerating Universe in modified gravity with dynamical torsion,''
  arXiv:1606.02565 [hep-th].








\bibitem{fT1}
J.~B.~Dent, S.~Dutta and E.~N.~Saridakis,
  ``f(T) gravity mimicking dynamical dark energy. Background and perturbation analysis,''
  JCAP {\bf 1101}, 009 (2011)
  doi:10.1088/1475-7516/2011/01/009
  [arXiv:1010.2215 [astro-ph.CO]].
  
  
\bibitem{fT2}
  S.~Nesseris, S.~Basilakos, E.~N.~Saridakis and L.~Perivolaropoulos,
  ``Viable $f(T)$ models are practically indistinguishable from $\Lambda$CDM,''
  Phys.\ Rev.\ D {\bf 88}, 103010 (2013)
  doi:10.1103/PhysRevD.88.103010
  [arXiv:1308.6142 [astro-ph.CO]].
  
  

\bibitem{fT3}  
G.~Kofinas and E.~N.~Saridakis,
  ``Cosmological applications of $F(T,T_G)$ gravity,''
  Phys.\ Rev.\ D {\bf 90}, 084045 (2014)
  doi:10.1103/PhysRevD.90.084045
  [arXiv:1408.0107 [gr-qc]].  



\bibitem{fT4}
Y.~F.~Cai, S.~Capozziello, M.~De Laurentis and E.~N.~Saridakis,
  ``f(T) teleparallel gravity and cosmology,''
  Rept.\ Prog.\ Phys.\  {\bf 79}, no. 10, 106901 (2016)
  doi:10.1088/0034-4885/79/10/106901
  [arXiv:1511.07586 [gr-qc]].
  
  \bibitem{fT5}
  R.~C.~Nunes, A.~Bonilla, S.~Pan and E.~N.~Saridakis,
  ``Observational Constraints on $f(T)$ gravity from varying fundamental constants,''
  Eur.\ Phys.\ J.\ C {\bf 77} (2017) no.4,  230
  doi:10.1140/epjc/s10052-017-4798-5
  [arXiv:1608.01960 [gr-qc]].
 


  
  




\bibitem{45}
V.~P.~Nair, S.~Randjbar-Daemi and V.~Rubakov,
  ``Massive Spin-2 fields of Geometric Origin in Curved Spacetimes,''
  Phys.\ Rev.\ D {\bf 80}, 104031 (2009)
  doi:10.1103/PhysRevD.80.104031
  [arXiv:0811.3781 [hep-th]].

\bibitem{33}
V.~Nikiforova, S.~Randjbar-Daemi and V.~Rubakov,
  ``Infrared Modified Gravity with Dynamical Torsion,''
  Phys.\ Rev.\ D {\bf 80}, 124050 (2009)
  doi:10.1103/PhysRevD.80.124050
  [arXiv:0905.3732 [hep-th]].


\bibitem{34}
C.~Deffayet and S.~Randjbar-Daemi,
  ``Non linear Fierz-Pauli theory from torsion and bigravity,''
  Phys.\ Rev.\ D {\bf 84}, 044053 (2011)
  doi:10.1103/PhysRevD.84.044053
  [arXiv:1103.2671 [hep-th]].






  
  \bibitem{BD}
  D.~G.~Boulware and S.~Deser,
  ``Can gravitation have a finite range?,''
  Phys.\ Rev.\ D {\bf 6} (1972) 3368.
  
  

\bibitem{dRGT} 
  C.~de Rham and G.~Gabadadze,
 ``Generalization of the Fierz-Pauli Action,''
  Phys.\ Rev.\ D {\bf 82}, 044020 (2010)
  doi:10.1103/PhysRevD.82.044020
  [arXiv:1007.0443 [hep-th]].
  
  
\bibitem{no_ghosts-1}
S.~F.~Hassan and R.~A.~Rosen,
  ``Resolving the Ghost Problem in non-Linear Massive Gravity,''
  Phys.\ Rev.\ Lett.\  {\bf 108}, 041101 (2012)
  doi:10.1103/PhysRevLett.108.041101
  [arXiv:1106.3344 [hep-th]].


\bibitem{no_ghosts-2}
S.~F.~Hassan and R.~A.~Rosen,
  ``Confirmation of the Secondary Constraint and Absence of Ghost in Massive Gravity and Bimetric Gravity,''
  JHEP {\bf 1204}, 123 (2012)
  doi:10.1007/JHEP04(2012)123
  [arXiv:1111.2070 [hep-th]].  

\bibitem{code}
\url{https://github.com/nikiforovavas/the_perturbations.git}

\end{thebibliography}
\end{document}